\documentclass[a4paper]{article}%[a4paper,twocolumn,10pt]

\usepackage{amssymb}
\usepackage{graphicx}
\usepackage{rotating}
\usepackage{amsmath}
\usepackage{mathtools}
\usepackage{times}
\usepackage{bm}
\usepackage{url}
\usepackage{multirow}
\usepackage{tabularx, ragged2e, booktabs}
\usepackage{xr}
\newcommand{\bfa }{} 
\newcommand{\bfb }{}
\newcommand{\bfc }{}
\newcommand{\bfd }{}
\newcommand{\bfe }{}
\newcommand{\bff }{}

\newcommand{\bfcacc }{ } % mind the whitespaces
\newcommand{\bfg }{}

\newcommand{\bfh }{}

\DeclareMathOperator\arctanh{arctanh}

\begin{document}

\title{
%Nonlinear forced change implies nonergodicity and vice-versa: The case of ENSO teleconnections
Nonlinear forced change and nonergodicity: The case of ENSO-Indian monsoon and global precipitation teleconnections} 

\author{\large Tam\'as B\'odai, G\'abor Dr\'otos, %Axel Timmermann, 
Kyung-Ja Ha, June-Yi Lee, \\ T\'imea Haszpra and Eui-Seok Chung \\
} 

\maketitle

\begin{abstract}

We study the forced response of the teleconnection between the El Niño–Southern Oscillation (ENSO) and global precipitation in general and the Indian summer monsoon (IM) in particular in the Max Planck Institute Grand Ensemble. The forced response of the teleconnection is defined as the time-dependence of a correlation coefficient evaluated over the ensemble. The ensemble-wise variability is taken either wrt. spatial averages or dominant spatial modes in the sense of Maximal Covariance Analysis or Canonical Correlation Analysis or EOF analysis. We find that the strengthening of the ENSO-IM teleconnection is robustly or consistently % 19.05.20. added after i call robustness into question wrt. the different origin of the change, ENSO variability versus coupling strength
featured in view of all four teleconnection representations, whether sea surface temperature (SST) or sea level pressure (SLP) is used to characterise ENSO, and both in the historical period and under the RCP8.5 forcing scenario. The main contributor to this strengthening in terms of a linear regression model is the regression coefficient, which can outcompete even a declining ENSO variability in view of using the SLP. We also find that the forced change of the teleconnection is typically nonlinear by (1) formally rejecting the hypothesis that ergodicity holds, i.e., that expected values of temporal correlation coefficients with respect to the ensemble equal the ensemble-wise correlation coefficient itself, and also showing that (2) the trivial contributions of the forced changes of e.g. the mean SST and/or precipitation to temporal correlations are insignificant here. We also provide, in terms of the test statistics, global maps of the degree of nonlinearity/nonergodicity of the forced change of the teleconnection between local precipitation and ENSO. 

\end{abstract}

\section{Introduction}\label{sec:Intro}

ENSO teleconnections are widely studied but their forced changes remain to be further explored and understood. Power and Delage~\cite{PD:2018} {\bfcacc provide} a multi-model assessment of ENSO-precipitation teleconnection changes based on the CMIP5 archive. They {\bfcacc consider}, in particular, ENSO-driven precipitation anomalies in tropical regions around the globe, and assess {them {\em jointly} with changes of mean precipitation}. For example, they find that, as a combination, wet DJF anomalies are {\bfg ``strengthened''} in the Tibetan plateau both by increasing mean precipitation and wet ENSO teleconnection strength; or, as another type of combination, an increasing JJA dry ENSO teleconnection strength is counteracted by a wetter mean state in Southeast Asia. Haszpra et al.~\cite{HHB:2020} {\bfcacc have} evaluated only the trend in ENSO-precipitation teleconnection strength changes, however, not in a model ensemble but the so-called ``single model initial condition (large) ensemble'' (SMILE) CESM1-LE~\cite{Kay:2015}. Working with a SMILE has the advantages that the forced response is correctly represented~\cite{BT:2012,DBT:2015,Tel:2019} in that model at least, and seeking a physical interpretation of changes is not faced with confusion at the outset, even if the physics is misrepresented by that one model.

The forced response is the time evolution of some ensemble-wise statistics, or, most generically, that of the probability measure carried by the climate {\em snapshot attractor}~\cite{BT:2012,DBT:2015,Tel:2019}. The ensemble-wise statistics evaluated over the converged ensemble~\cite{DBT:2017} is in a one-to-one correspondence with the external forcing of the nonautonomous dynamical system, hence the term `forced response'. Without convergence, or considering a single realisation, {\bfc which may even correspond to} observations, the state depends also on initial conditions, {\bfcacc so that} forced changes cannot be precisely disentangled from changes brought about by internal variability. In an attempt of doing this nevertheless, concerning single realisations, the standard practice is that a trend, linear or not~\cite{Franzke2014}, is simply identified as a forced change, before possibly ``anomalies'', i.e., (what is assumed to be) the internal variability is analysed. Identifying the principal component (PC) of the first empirical orthogonal function (EOF1)~\cite{storch_zwiers_1999} obtained {\em without detrending} with the forced response~\cite{Kim:2015} is still somewhat arbitrary. It certainly leads to biases~\cite{DBT:2016}. Yet, even with SMILEs available, it is very common to see that authors evaluate temporal statistics first -- unnecessarily involving a subjective factor -- and ensemble statistics afterwards~\cite{VW:2017,Carreric:2020}. Otherwise, the biases are certainly controlled by the magnitude of the climate change signal relative to the intensity of internal variability -- a kind of signal-to-noise ratio. Computing the PC1 of EOF1 without detrending, as mentioned just above, is in fact meant to maximise the signal at least, and the EOF1 is referred to~\cite{Santer:2019} as a ``fingerprint'' of the forced change -- the spatial pattern that is supposed to be distorted by internal variability, in any single realisation, the least. See~\cite{Timmermann:1999} that is concerned rather with the minimisation of the ``noise''.

It is not pertinent to talk about ``advantages'' of temporal or ensemble methods over one another, because no situation arises when we need to decide between them. When an ensemble is {\em available}, the correct, conceptually sound ensemble method is to be applied if the forced response {\bfg(including that of the internal variability)} is concerned. However, certainly we can make statements only about {\bfcacc the given} model due to model errors. Therefore, we should speak only about {\em limitations} in the {\em respective} situations of analysing observational or model ensemble data. 

Concerning teleconnections, in particular, the forced response can be identified by the time evolution of the ensemble-wise Pearson correlation coefficient~\cite{Herein:2016,Yettella:2018} (in a simple linear approach) between some anomaly quantities. The anomaly can {\bfc correspond to} simply a spatial (mean of a temporal) mean~\cite{BDHLL:2020}, but also the PC of an EOF concerning variability {\em across the ensemble}, called a snapshot EOF (SEOF)~\cite{HTH:2020}. Haszpra et al.~\cite{HHB:2020} {\bfcacc take} the latter approach; however, it can be extended to obtaining anomalies observing the ``mutual variability'', such as in the sense of Maximal Covariance Analysis (MCA)~\cite{storch_zwiers_1999} or Canonical Correlation Analysis (CCA)~\cite{Hardle2007,storch_zwiers_1999}. {\bfh With an interest in a teleconnection and its forced response}, MCA and CCA -- just like EOF analysis -- can also be pursued concerning the variability across the ensemble, whereby we can refer to these methods as SMCA and SCCA. This is perhaps best suited to teleconnection analyses concerning two extended, possibly disconnected, regions. See an application of SMCA to studying the forced change of the coupling of JJA 200-hPa geopotential height and September sea ice concentration in~\cite{Timi_Variations}.  

An important example of this would be the relationship of the summer monsoon precipitation on the Indian subcontinent with the ENSO phenomenon {\bfc which extends over the whole} of the Equatorial Pacific~\cite{bin_wang,Mishra:2012,Wang:2015}. In a previous publication~\cite{BDHLL:2020} we examined the forced response of the ENSO-Indian monsoon teleconnection in the Max Planck Institute Grand Ensemble (MPI-GE)~\cite{Maher:2019} representing the monsoon by the average JJAS precipitation over India and the ENSO by either the gridpoint- and SLP-based SOI, or the areal mean of the SST in some extended area in the Equatorial Pacific. Standard choices for the latter are the Ni\~no3, Ni\~no4, and Ni\~no3.4 regions. For the first time we established, via a formal hypothesis test, that standard representations of the teleconnection in the sense of a correlation coefficient between some of the above-mentioned indices {\em strengthened} in this model, both in the historical period and under the strong RCP8.5 forcing scenario, although not necessarily monotonically. In fact, using the SOI it was the historical period only when an increase could be detected, and, using the Ni\~no3 index, it was rather the RCP8.5 scenario under which an increase could be detected. This raises the question whether the discrepancy is down more to (1) the physical difference between the SOI and Ni\~no3 indices, or (2) rather that -- one based on two distant gridpoints while the other on a limited region of the Equatorial Pacific -- they ``take somewhat different slices'' of the whole ENSO phenomenon, which can matter when even spatial characteristics of ENSO have a forced response beside possibly the ``temporal''. In fact, using the Ni\~no3.4 index, instead of Ni\~no3, the result is {\bfc closer to} that using the SOI, or, even more so using a box-SOI, even to the extent of detecting the same {\em nonmonotonicity}: a temporary weakening around the turn of the millennium. 
In addition, both with Ni\~no3.4 and the box-SOI we can detect changes {\em both} under the historical and scenario forcing. On the one hand (I), this seems to rule out that using SST versus SLP makes much difference. We will see below that this is actually not precisely the case. On the other hand (II), it is still a question whether spatial characteristics of ENSO, in terms of SEOFs, or the whole teleconnection, in terms of SMCA or SCCA, would also change, or only the variance as {\bfc reflected in the magnitudes of} the PCs belonging to either of the said instantaneous/snapshot spatial modes. These are the two basic questions that we set out here to investigate.

Furthermore, {\bfb this study identifies, in terms of a regression model, controlling factors driving the forced change of the ENSO-IM teleconnection strength as being}: the ENSO variability, the coupling (regression coefficient), and the intensity of other influences -- {\bfc which} can be viewed as noise. We find that the changes of the ENSO-IM teleconnection are not driven only or dominantly by {\bfb the change of} ENSO variability. {\bfb Nevertheless, it imprints its nonlinearity onto the teleconnection change, which could be an important source of biases as follows.} We observe in the MPI-GE a nonmonotonic change in the ENSO variability: after a seemingly monotonic change, a decline follows in the second half of the 21. c. under RCP8.5. Such a feature is absent in the CESM1{\bfg -LE}~\cite{HHB:2020}. This nonmonotonicity and possibly others, as already mentioned above, constituting a nonlinear change, prompt that the system should be nonergodic also with respect to correlations, {\bfe i.e.,} biases should exist~\cite{DBT:2016} in the temporal correlation coefficient evaluated in, say, multi-decadal time windows. Such a bias is something to bear in mind beside the ample statistical fluctuations of finite-size statistics even under a stationary climate~\cite{BDHLL:2020}. Given that the ensemble size is finite and not so large concerning teleconnections, we develop here a statistical test whereby we can detect nonergodicity, and, subsequently, map out regions of the world where such a nonergodicity can be detected in the MPI-GE in the context of the relationship of local precipitation with ENSO. 

The rest of the paper is organised as follows. In Sec.~\ref{sec:methods} we provide details about our methodology of analysing the forced response of teleconnections, such as the way we pursue EOF, MCA, CCA, as well as the idea of decomposing the change of the teleconnection strength in terms of a simple regression model. In Sec.~\ref{sec:results} we provide results both on spatial and temporal aspects of the forced change of the ENSO-IM teleconnections. In Sec.~\ref{sec:nonergodicity} we outline our method of detecting nonergodicity and map out the world with respect to the degree of nonergodicity concerning the synoptic relationship of the local JJA precipitation with ENSO. In Sec.~\ref{sec:discussion} we discuss and summarise our results. 

\section{Data and methods}\label{sec:methods}

\subsection{Data}

The analysis is restricted to the use of the MPI-GE data~\cite{Maher:2019}, that is, to a particular Earth system model (ESM), the MPI-ESM. The quoted presentation paper of the data set outlines main features of the state-of-the-art MPI-ESM. To characterise the ENSO we make use of the 2D fields of either the Sea Level Pressure (SLP) or the Sea Surface Temperature (SST). The Indian summer monsoon rain (ISMR) was calculated from the total (convective and large scale) precipitation rate. The variable codes for SLP, SST and total precipitation rate are 134, 12 and 4, respectively.
{\bfh Monthly mean SLP is accessible publicly at \url{https://esgf-data.dkrz.de/projects/mpi-ge/} under variable name `psl'.} The SST can be derived from the top layer of the 3D potential temperature field, whose variable is `thetao'. Instead of the total precipitation rate [km/s], one can use the precipitation flux [kg/s/m$^2$], whose variable name is `pr'. 

\subsection{Representations of the ENSO-IM teleconnection}

The scalar quantity of the ISMR averaged in time over the JJAS monsoon season and in space over India is the same as that calculated in~\cite{BDHLL:2020}, in order to secure a correspondence with the so-called AISMR rain gauge data set~\cite{Parthasarathy:1994} (excluding some states of India; see their Fig. 1). Scalar quantities to represent ENSO variability are also the same as in~\cite{BDHLL:2020}, as follows. 1. Average SST in the standard Ni\~no3 region represented by the box (5$^{\circ}$S, 5$^{\circ}$N, 210$^{\circ}$E, 270$^{\circ}$E); 2. Average SST in the standard Ni\~no3.4 region (5$^{\circ}$S, 5$^{\circ}$N, 190$^{\circ}$E, 240$^{\circ}$E); 3. SLP difference, $p_{\rm diff} = p_{\rm Tahiti} - p_{\rm Darwin}$, between the closest gridpoints to Tahiti and Darwin. 4. The difference of the average SLPs in the boxes (5$^{\circ}$S, 5$^{\circ}$N, 80$^{\circ}$E, 160$^{\circ}$E) and (5$^{\circ}$S, 5$^{\circ}$N, 200$^{\circ}$E, 280$^{\circ}$E). We will refer to the average SSTs under 1. and 2. as the Ni\~no3 and Ni\~no3.4 indices, respectively, and to the pressure differences under 3. and 4. as the SOI and box-SOI indices (SOI being short for the Southern Oscillation Index), respectively. See a discussion on this in Sec. 4.a of~\cite{BDHLL:2020}. 

We correlate the ``near-synoptic'' JJA mean of {\bfc any of} Ni\~no3, Ni\~no3.4, SOI, box-SOI with the JJAS (model) AISMR. The correlation coefficient $r$ is evaluated {\em across the ensemble}, as done in~\cite{BDHLL:2020}, which results in annual time series. These time series provide {\em representations} of the forced response of the ENSO-IM teleconnection. However, the finite ensemble size entails large statistical error fluctuations, which badly mask the forced response signal. That is, the estimate $\hat{r}(t)$ rather poorly represents the unknown true $r(t)$ signal. Therefore, we aspire only to {\em detect a change} in certain time periods. For this, we employ the Mann-Kendall test~\cite{Mann:1945} in the same way as done in~\cite{BDHLL:2020}, aiming to reject the hypothesis of stationarity against the alternative hypothesis of a monotonic trend masked by serially uncorrelated noise constituted by realisations of iid rv's (independent identically distributed random variables). 

We will evaluate correlations of the AISMR also with the PCs {\bfc (which are scalars)} belonging to EOFs. PCs and EOFs can be also composed with respect to (wrt.) the ensemble, as a ``snapshot'' (never mind the seasonal means), hence the name SEOF, as proposed by~\cite{HHB:2020,HTH:2020}\footnote{Ensemble-wise EOFs, SEOFs, were used earlier in the context of probabilistic ensemble forecasting in e.g.~
\cite{doi:10.1175/WAF-D-12-00132.1,doi:10.1175/WAF-D-16-0112.1,doi:10.1175/MWR-D-18-0052.1}, in which case the concern was certainly not the forced response.}. The SEOFs are evaluated here wrt. the Equatorial Pacific box (30$^{\circ}$S, 30$^{\circ}$N, 150$^{\circ}$E, 295$^{\circ}$E). {\bfc Traditional EOFs} are decomposing spatio-temporal variability as a sum of standing {\bfc waves, i.e., time-independent spatial patterns} modulated by arbitrary {\bfc (but uncorrelated)} temporal signals, whereby the spatial {\bfc patterns are} given by the EOFs, and the temporal signals by the PCs. {\bfc Snapshot EOFs are the same but different time steps are replaced by different ensemble members (i.e., realizations of the dynamics).} EOFs are orthogonal, and each new EOF is defined such that the corresponding PC has the largest possible variance, leading to an eigen-problem. The variances of the PCs are singular values belonging to the EOFs being the singular vectors. We find them by using Matlab's \verb|svd| as done in~\cite{eofmanual}. The ordering of the singular values wrt. magnitude provides a natural ordering of the EOFs, which are denoted as EOF1, EOF2, etc., and like-wise we write PC1, PC2, etc. Correlating e.g. the PC1s of either the SLP or the SST field {\bfg(in the same box)} with AISMR {\bfg have more in common than the representations given by e.g. the pairs of Ni\~no3-AISMR and SOI-AISMR, because Ni\~no3 and SOI do not derive from the same mathematical definition of dominance~\footnote{\label{fn:rationale}{\bfg Although}, the rationale behind defining {\bfc simple} indices is such that we want to work with a simple-to-compute quantity {\bfc (e.g. an areal average or difference between two locations) that correlates} very strongly with the dominant variability in terms of EOFs, which latter can be seen {\bfc to represent} a more natural method of decomposing the variability of a field.}; \bfg and one could think that more similar representations yield more similar results. Therefore}, firstly, we wish to see if we can establish a {\em robust} representation of the ENSO-IM teleconnection by having possibly a better match {\bfc between} the forced response signals belonging to the SLP and SST than between those using the classic indices 1.-4. Secondly, the teleconnection strength can change possibly because of a shift in the ``centre of action'' of the ENSO phenomenon, something that can impact on the teleconnection representations by 1.-4. {\em differently}. Thirdly, {\bfc including} higher EOF modes can reveal more predictive power of the full e.g. SST field compared with e.g. Ni\~no3 alone~\footnote{Given that the PCs are uncorrelated, one can calculate the so-called coefficient of multiple correlation or determination~\cite{storch_zwiers_1999} as the square root of the squared correlation coefficients between the ``predictand'' and the individual PCs, which provides a framework for identifying the {\em contribution} of the different PCs to predictability in terms of a multi-dimensional linear regression model.}. In this regard we are interested in whether the correlations to do with the higher modes are more susceptible to the anthropogenic forcing. We will only show results for the second modes.

We can pursue the three points of inquiry above also in terms of MCA and CCA. {\bfc MCA and CCA are similar to the EOF analysis, but consider {\em two} fields, and find the orthogonal modes for which the covariance and the correlation, respectively, is maximal between the corresponding PCs of the two fields.} We carry out these analyses by applying Matlab's \verb|svd| and \verb|canoncorr|, respectively. These methods will allow us to take into account a change in spatial characteristics on the side of the IM too. We will use the box of (5$^{\circ}$S, 40$^{\circ}$N, 65$^{\circ}$E, 100$^{\circ}$E) on the side of the IM. We evaluate the correlation coefficient between the PC1s of the respective 1st SMCA/SCCA modes as well as the PC2s. That is, the correlations of these PC2s are not analogous to those of the AISMR and PC2 of EOF2. We anticipate that the SMCA and SCCA lead to higher correlations than the SEOF analysis, stemming from their very definition. In this regard, as the fourth point of inquiry, we want to see if higher correlations are more susceptible to anthropogenic forcing. {\bfc Instead of the complete field, the SCCA is performed on the PCs corresponding to the first 4 {\bfh SEOFs, since it is ill-defined on the full fields.}} We will speak about SCCA modes as the linear combination of the SEOF modes in terms of the coefficients defining the so-called canonical variables. Indeed, thanks to the orthogonality of the SEOFs, the projection of the full fields onto these modes yield the SCCA PCs whose correlations were maximised by the SCCA.

\subsection{Decomposition of the forced change of the teleconnection strength}\label{sec:decomp}

Instead of trying to find physical mechanisms responsible for a particular forced change signal {\bfc in} the teleconnection strength, here we pursue only a statistical attribution of that. In terms of a linear regression model, $\Psi = a\Phi + \xi$, underpinning the correlation coefficient, $r$, we can attribute {\bfc changes in the teleconnection strength} to three factors. We can identify these factors {\bfc by considering} the formula:
\begin{equation}\label{eq:r}
 r=\frac{1}{\sqrt{1+(\frac{\sigma_{\xi}/a}{\sigma_{\Phi}})^2}}.
\end{equation}
These factors are:
\begin{itemize}
 \item ENSO variability $\sigma_{\Phi} = \textrm{std}[\Phi]$;
 \item ENSO-IM ``coupling'' $a$ being the regression coefficient;
 \item Noise strength $\sigma_{\xi} = \textrm{std}[\xi]$.
\end{itemize}
Note that, like $r$, all {\bfc $\sigma_{\Phi}$ and $\sigma_{\xi}$} are defined in terms of the variability wrt. the ensemble, for every year separately. That is, like $r(t)$, we have time series {\bfc $\sigma_{\Phi}(t)$ and $\sigma_{\xi}(t)$, and also $a(t)$}. Based on these, in a simple way, we can say that e.g. a change in $r(t)$ in a time period is due to a change in $\sigma_{\Phi}(t)$ if in that period no change is seen wrt.  $a(t)$ and $\sigma_{\xi}(t)$. Or, perhaps we can also say that if $\sigma_{\xi}/a$ shows no change. In fact, it has been suggested to us that perhaps the strengthening of the ENSO-IM teleconnection is due in this model to an increasing ENSO variability. Although apriori this possibility is perhaps hard to exclude, we note that it would be rather specific, and could possibly be seen accidental, to the ENSO-IM teleconnection given that ENSO-precipitation teleconnections can either strengthen or weaken in different places on Earth~\cite{PD:2018}. This is shown in Fig. 5 of~\cite{HHB:2020} for the CESM1-LE; and the MPI-GE data shows a rather similar picture as seen in our Fig.~S1 %\ref{fig:S09} 
in the Supplementary Material. 

In order to check the hypothesis of the dominant role of ENSO variability, one can evaluate $\sigma_{\Phi}$ directly {\bfc (i.e., compute the standard deviation of $\Phi$ over the ensemble)}, on the one hand, and $\sigma_{\xi}/a$ by simply inverting Eq.~(\ref{eq:r}), on the other. {\bfa (The only issue with the latter is that the sign of $a$ remains indeterminate.)} Anticipating that $\sigma_{\xi}/a$ is not constant in time, we can evaluate $a$ by, first, directly evaluating the IM variability $\sigma_{\Psi}$, and, subsequently, using the textbook formula: $$a=r\frac{\sigma_{\Psi}}{\sigma_{\Phi}}.$$ (We can, of course, evaluate $a$ directly by linear regression, but calculating the standard deviation is easier, and we already have $r$ on hand.) In turn, having now also $a$ on hand, $\sigma_{\xi}$ can be obtained by inverting Eq.~(\ref{eq:r}).

{\bfa From eq. (\ref{eq:r}) we see {\bfc the following: if} in a time period $[t_1,t_2]$ the ENSO variability increases as $\sigma_{\Phi}(t_2)=\beta\sigma_{\Phi}(t_1)$, $\beta>1$, and we have also a decrease $\sigma_{\xi}(t_2)/a(t_2)=\alpha\sigma_{\xi}(t_1)/a(t_1)$, $\alpha<1$, then $\alpha\beta>1$ would mean that the ENSO variability increase, in comparison with the $\sigma_{\xi}/a$ decrease, has a larger effect on the increase of $r(t)$ in that period. Given the very noisy time series, the appropriate approach would be a statistical test attempting to reject the null-hypothesis of $\alpha\beta=1$. However, it is not clear to us how this can be done. As for a preliminary analysis, we simply evaluate $\alpha\beta$ in all possible time periods $[t_1,t_2]$ assuming linear changes of $\sigma_{\Phi}(t)$ and $\sigma_{\xi}(t)/a(t)$. That is, $\alpha$ and $\beta$ are determined from the respective linear regression model fits.}

\section{The forced response of the ENSO-IM teleconnection}\label{sec:results}

\subsection{Spatial aspects}\label{subsec:spatial}

We start by inspecting the forced response of spatial characteristics of the ENSO and IM variability and those of their teleconnection. Fig.~\ref{fig:01} shows the first modes of the SEOF, SMCA, SCCA analyses on the side of the ENSO, based on the SST, {\bfc smoothed by averaging within} four subsequent 50-year periods starting from 1900. The temporal averaging brings it really out that hardly any change is visible {\bfc in the EOF1 and the MCA1 mode,} even upon the strong RCP8.5 forcing. Carreric et al.~\cite{Carreric:2020} suggests that the maximum of the EOF1 mode shifts to the east in the CESM1-LE by showing that some centre (C) mode~\cite{Takahashi:2011} shifts to the east more in comparison with the westward shift of some east (E) mode~\footnote{This situation is not shown conclusively in~\cite{HHB:2020} because only three snapshots are shown belonging to single years, and it is not clear how much statistical error is associated with them, leaving an uncertainty about the locations of the peaks.}. {\bfc This does not appear to be so here, although the error of the results (with respect to any year within a given period) would be difficult to be estimated.} % ***Mention the eastward shift in CESM2, already in the historical period? 
{\bfc Unlike the EOF1 and the MCA1 mode (which are very similar to each other), the CCA1 mode does undergo a considerable change by the late 21st century. It is even more interesting to observe this change to have a different sign in the middle of the ENSO domain (highlighted by the Ni\~no3 and Ni\~no3.4 boxes) than the minor change in the EOF1 and the MCA1 mode. %, as shown in Fig. S***. 
Since the MCA1 mode is supposed to mainly reflect the ENSO variability (see later), unlike the CCA1 mode, which truly captures the variability relevant for the teleconnection, this discrepancy presumably means (taking into account the patterns in the reference period) that important areas for ENSO itself and for its teleconnection with the Indian monsoon get more decoupled. In particular, the middle of the basin gains importance for ENSO but loses importance for the teleconnection.} The supplementary figure Fig.~S2 %\ref{fig:S03} 
{\bfc suggests} the same picture for the SLP.

{\bfc Although each panel of Fig.~\ref{fig:01} washes together 50 separate entities, the mostly monotonic emergence of the changes in consecutive rows suggests that we can see meaningful signals.} Otherwise, the supplementary video \url{https://youtu.be/RdHjbjZxaZQ} shows ample year-to-year change in the form of a flickering. Given that the forcing is rather smooth and gradual, this is clearly just a finite-ensemble-size statistical error fluctuation.

Interestingly, the second modes of ENSO variability, both in terms of SST (Fig.~S3) %\ref{fig:S01}) 
and SLP (Fig.~S4), %\ref{fig:S04}), 
are largely unchanged, too. We note, however, that the obtained temporal-mean second modes are less reliable than the first modes for the following reason. {\bfc The sign of a mode is arbitrary, and we need to choose a sign convention, according to which we flip the modes returned by the \verb|svd| Matlab routine in years when needed.} To this end we pick a reference year, and calculate the spatial correlation of the modes in all years with the mode in the reference year. When the sign of the correlation is negative, then we flip the sign/orientation of the mode. However, when the statistical error corrupting the modes is too large, the pattern matching might not be possible. It seems to be the case that the pattern matching is feasible for the first modes, but fails in certain years for the second modes. See the supplementary videos \url{https://youtu.be/bmbPXAtq_p4} and \url{https://youtu.be/4pWPf2oUE-I}. 

We can achieve a smoothing of the temporal evolution of the modes by pooling {\em snapshot} (annual) anomalies from a temporal window for the evaluation of the modes. That is, we mix temporal and ensemble-wise variability. We emphasize, however, {\bfc that} {\bfc the} anomalies are obtained the same way as done without smoothing: simply by subtracting the ensemble-mean belonging to {\em single} {\bfc years, so that the only subjective element is the choice of the window length}. The supplementary videos (\url{https://youtu.be/bmbPXAtq_p4} \url{https://youtu.be/R78ycuvOI6c} \url{https://youtu.be/4pWPf2oUE-I} \url{https://youtu.be/RdHjbjZxaZQ}) show that the effect of smoothing by a 5-year window is well visible, but would still allow for modes to be far ``out of shape''. Considering the comparison of 50-year periods, however, it is not clear to us which way would smaller errors be incurred -- whether by averaging 50 annual ``snapshot'' modes or by pooling {\em anomaly} data from the full 50-year period -- because there is no reason to think that the finite-ensemble-size estimate is unbiased~\footnote{As an example, the sample correlation coefficient for normally distributed rv.s is negatively biased~\cite{Fisher:1915}}. It is certainly computationally far less intensive by pooling the data, because then the expensive SVD needs to be performed only once.

% \begin{figure}
%   \begin{center}
%       %\includegraphics[width=\linewidth]{ENSO_IM_teleco_12_mean_snapshots_ENSO_mode_1_w.png} 
%       \includegraphics[width=\linewidth]{ENSO_IM_teleco_12_mean_snapshots_ENSO_mode_1_diff.jpg} 
%       \caption{\label{fig:01} Forced change of the first modes of JJA-mean SST variability in the Equatorial Pacific. Temporal means are taken in four consecutive 50-year periods starting from 1900. The top row displays the temporal mean in the first period, and the subsequent rows display the difference with respect to that in the following periods. Left: EOF, middle: MCA, right: CCA first mode. For the MCA and CCA analysis the boxes on the side of the Indian monsoon are shown in Fig.~\ref{fig:02}. The Ni\~no3 and Ni\~no3.4 boxes are marked in gray and black, respectively.}
%   \end{center}
% \end{figure}

On the side of the IM, Fig.~\ref{fig:02} evidences a change in spatial characteristics, namely, a shift of some ``teleconnection hotspots'' in Northern India and over the sea to {\bfg North-Northwest and to North, respectively}, at least in the last 50 years. {\bfc Note that the EOFs, the MCA modes and the CCA modes, including their changes, are very similar. The biggest {\bfd dissimilarity} is again shown by the CCA mode as the relative weakness of a peak over the sea West of the land, but one may reasonably argue that the teleconnection with ENSO {\bfd (captured by CCA1 and MCA1)} is strongly bound to the strongest Indian variability {\bfd (captured by EOF1 and MCA1)}. In view of this, the coincidence of EOF1 and the MCA1 mode for ENSO may be attributed to the dominating variability of ENSO in covariance (as opposed to correlations, which determine the CCA1 mode). In Fig.~\ref{fig:02},} we show also the correlation {\bfc map and its changes} concerning the relationship of the gridpoint-wise precipitation with the PC1 of EOF1 {\bfc of ENSO, and find} similar patterns to those of the {\bfc first} SEOF/SMCA/SCCA {\bfh modes, which further supports the close relationship between the Indian precipitation variability and the ENSO-influence.} {\bfd Note that many areas of the strongest changes fall outside the AISMR region.} {\bfe In particular, most of the sea loses very much of its relevance for the IM variability and for its teleconnection with ENSO (a positive change results in getting closer to zero in these areas of negative sign).} 
Like the first modes, so do the second modes show changes in the last 50 years (Fig.~S6). %\ref{fig:S02}). 
{\bfc This further confirms that the teleconnection with the ENSO region is geographically rearranged over India.} Using SLP instead of SST on the side of the ENSO, the spatial patterns of precipitation over India are largely unchanged for both the first and second modes (Figs.~S5, S7). %\ref{fig:S05}, S\ref{fig:S06}). 
These features can be seen also in the supplementary videos \url{https://youtu.be/bmbPXAtq_p4} and \url{https://youtu.be/R78ycuvOI6c}. 

% \begin{figure}
%   \begin{center}
%       \includegraphics[width=1.\linewidth]{ENSO_IM_teleco_12_mean_snapshots_IM_mode_1_diff.jpg} 
%       \caption{\label{fig:02} Same as Fig.~\ref{fig:01} wrt. the first three columns, but concerning the Indian summer monsoon. In the fourth column a correlation map and its changes are shown concerning the gridpoint-wise JJAS precipitation and the PC1 of the EOF1 of the SST in the box seen in Fig.~\ref{fig:01}.}
%   \end{center}
% \end{figure}

As a curiosity, we {\bfc further elaborate on the similarity and the differences between different modes. As for the first ENSO modes, CCA1 gives prominence to the variability in the westernmost part of the selected box in comparison with EOF1 and MCA1}. Presumably this is in large part for the intuitive reason that for a maximal correlation (not covariance) it is what happens in the Pacific the closest to India that matters the most. For the SST, this comparison persists {\bfg to an extent} for the second EOF, MCA, CCA modes, {\bfg with somewhat less similarity of EOF2 and MCA2}. The SLP shows a similar behaviour. As for the IM, {\bfc MCA2 is an outlier, but matching is very good in all other cases}.  

\subsection{The forced evolution of correlation coefficients and the drivers of their change}
 
The time evolution of the correlation coefficients is given in the top row of Fig.~\ref{fig:03} for the SST-based representations of ENSO. This time evolution is basically a long-term increase for the Ni\~no indices (reproduced from~\cite{BDHLL:2020}) and for the first modes. The second modes behave differently: their correlation coefficients are practically constant, so that they really cannot have higher than secondary importance for changes in the teleconnection strength. 

For the Ni\~no indices and the first modes, the details of the detectable time evolution are presented in Fig.~\ref{fig:05} by temporal maps of the Mann--Kendall test statistic for stationarity, following~\cite{BDHLL:2020}. The general increase in teleconnection strength is detected by every quantifier of ENSO and the IM. Although a change is hardly detected within the 21st c. part alone (upper right triangle), the gradual darkening of the shades in the upper left box suggests a {\bff continued strengthening} of the teleconnection. {\bff Note that the radiative forcing is much stronger in the 21st c. than in the 20th c., so that any strengthening in the 21st c. must be nonlinearly slowed down in comparison with the 20th c.}

The further details also mostly match in all representations utilizing the AISMR (Figs.~\ref{fig:05}(a)-(c)), although an early-21st-c. drop does not surpass the threshold for 5\% significance for Ni\~no3, and the late-21st-c. increase is rather moderate for Ni\~no3.4. This feature is even more pronounced in the EOF1-EOF1 and the MCA1 representation (Figs.~\ref{fig:05}(d)-(e)), whereas the CCA1 representation comes with a very prominent gradual increase extending to all of the 21st c. (Fig.~\ref{fig:05}(f)). Interestingly, all representations based on principal modes on the IM side place the slight drop at the turn of the 20th-21st centuries considerably earlier than the AISMR-based representations.

\begin{figure}
  \begin{center}
      \includegraphics[width=\linewidth]{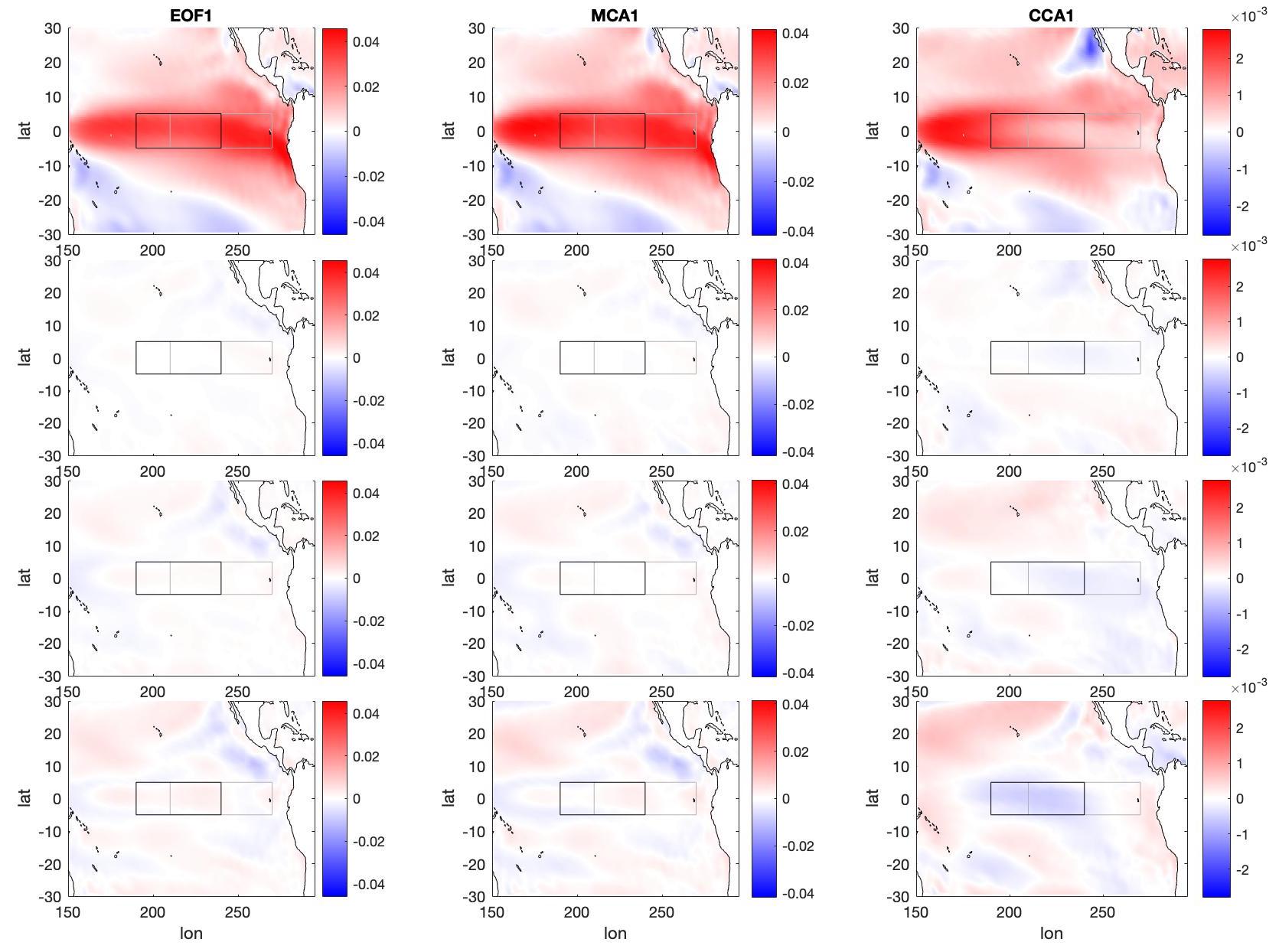} 
      \caption{\label{fig:01} Forced change of the first modes of JJA-mean SST variability in the Equatorial Pacific. Temporal means are taken in four consecutive 50-year periods starting from 1900. The top row displays the temporal mean in the first period, and the subsequent rows display the difference with respect to that in the following periods. Left: EOF, middle: MCA, right: CCA first mode. For the MCA and CCA analysis the boxes on the side of the Indian monsoon are shown in Fig.~\ref{fig:02}. The Ni\~no3 and Ni\~no3.4 boxes are marked in gray and black, respectively.}
  \end{center}
\end{figure}

\begin{figure}
  \begin{center}
      \includegraphics[width=1.\linewidth]{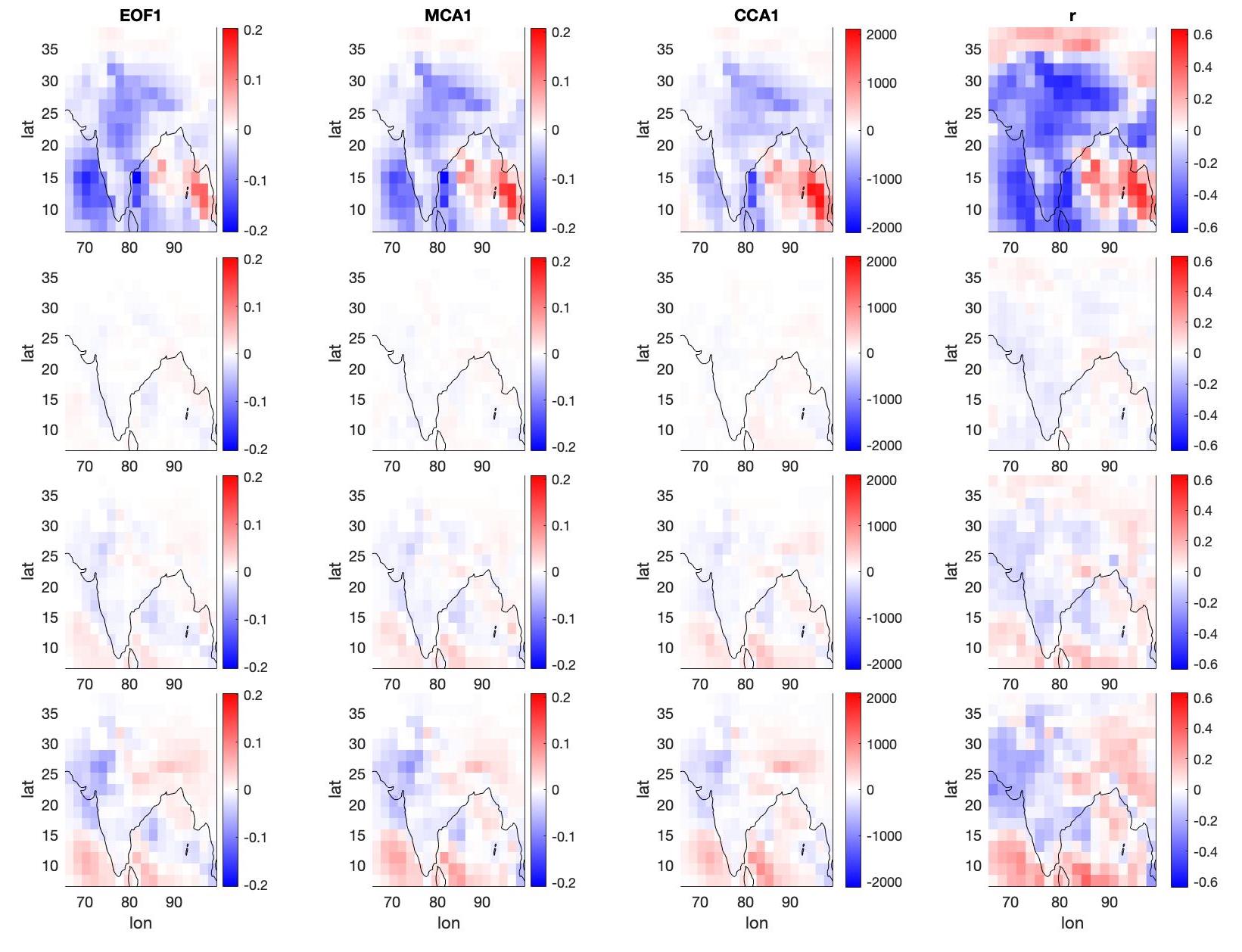} 
      \caption{\label{fig:02} Same as Fig.~\ref{fig:01} wrt. the first three columns, but concerning the Indian summer monsoon. In the fourth column a correlation map and its changes are shown concerning the gridpoint-wise JJAS precipitation and the PC1 of the EOF1 of the SST in the box seen in Fig.~\ref{fig:01}.}
  \end{center}
\end{figure}

\begin{figure}
  %\begin{center}
  %\centering
        \includegraphics[width=1.\linewidth]{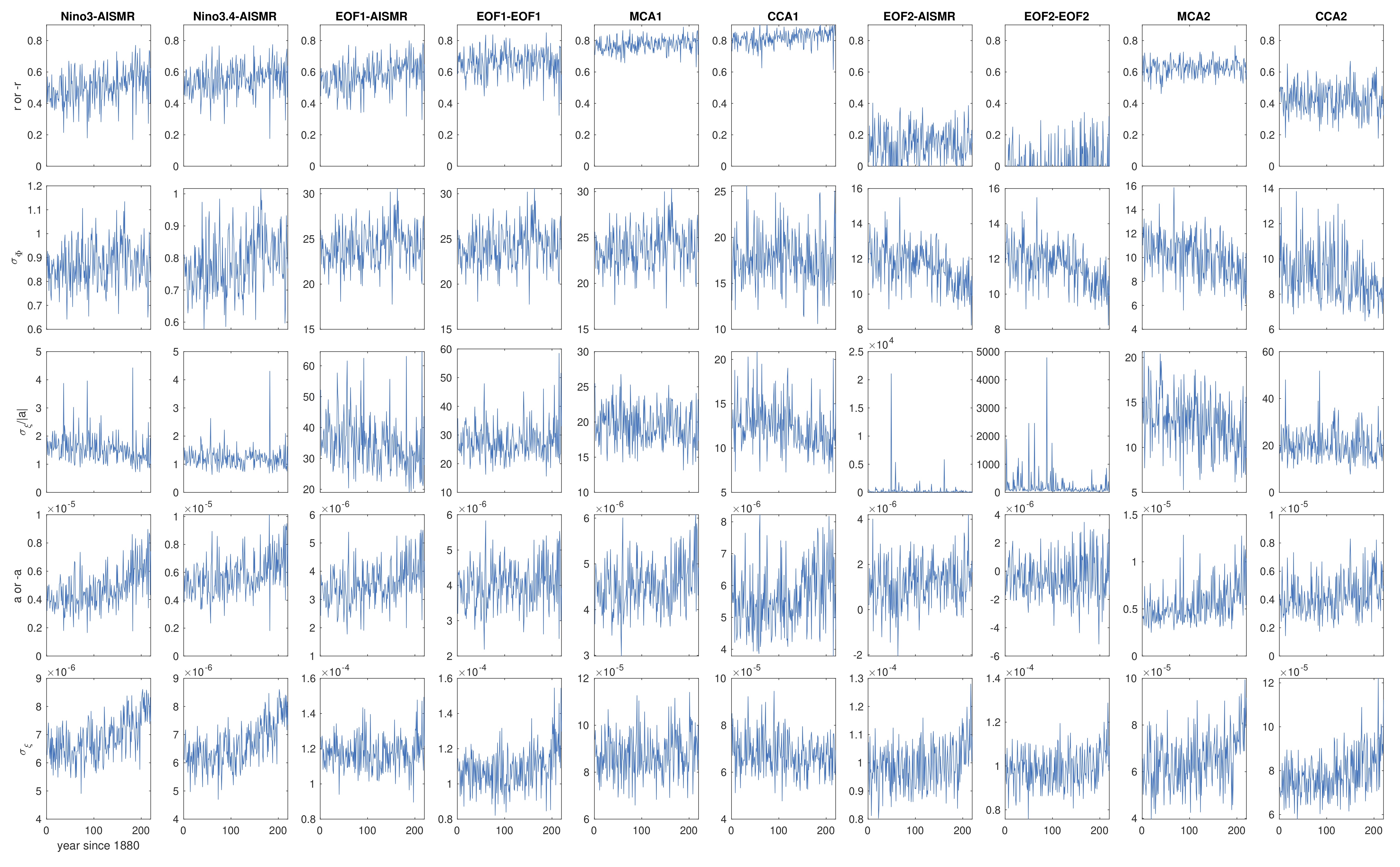}
        \caption{\label{fig:03} The forced evolution of correlation coefficients $r(t)$ and the drivers of change; see the main text. The different columns correspond to different representations of the ENSO-IM teleconnections. For columns 1-3, 7 the IM is represented by the average monsoon rain ISMR. Columns 1-6 concern the respective dominant modes of variability, and columns 7-10 concern the next-to-dominant modes of variability.}
    %\end{center}
%    \end{landscape}
 \end{figure}

\begin{figure}
  \begin{center}
            \includegraphics[width=\linewidth]{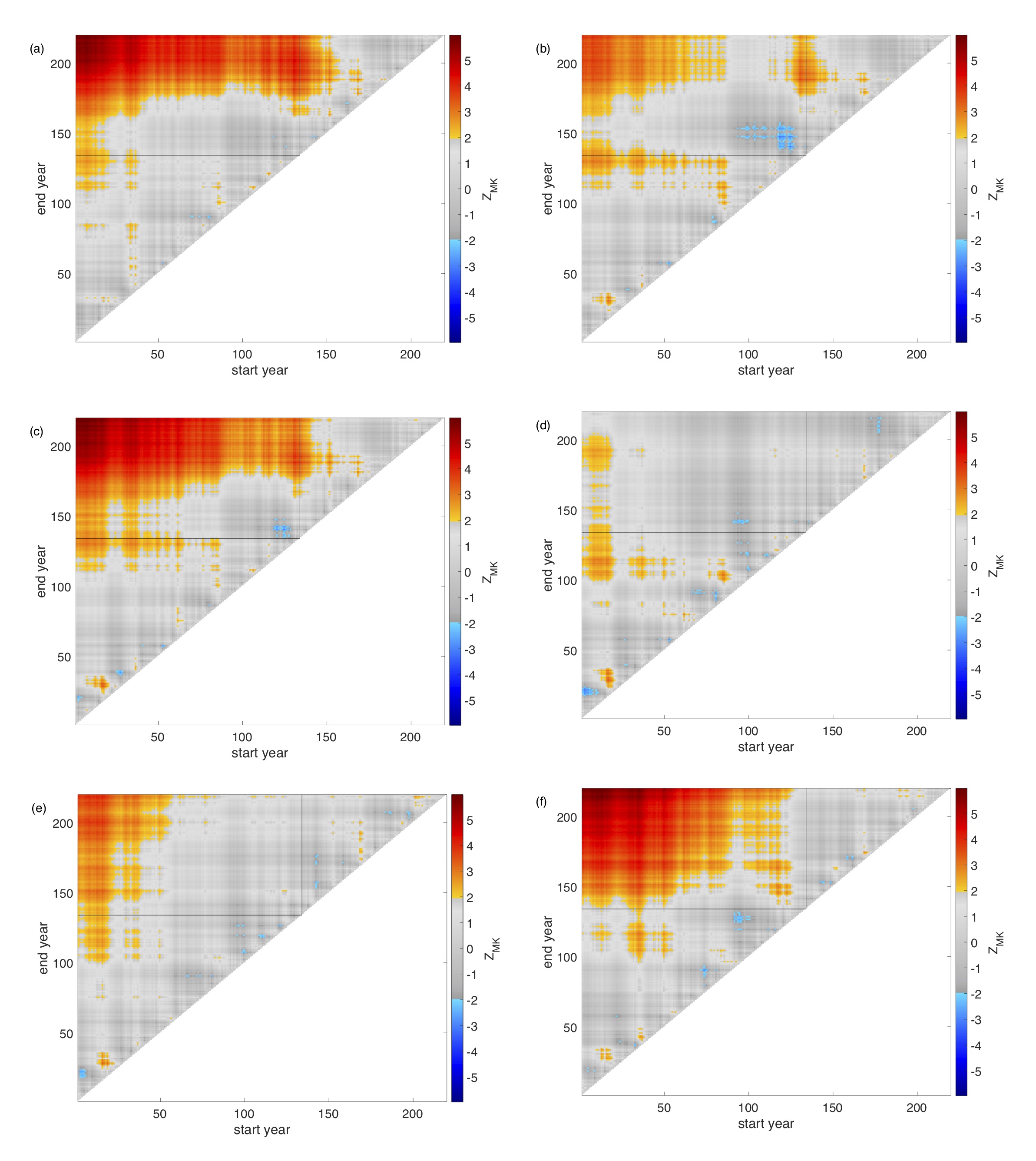} 
        \caption{\label{fig:05} Test statistics of the Mann-Kendall test for the stationarity of $r(t)$ {\bfa or $-r(t)$, depending on which of these gives an overall positive value}. {\bfg The diagrams correspond to columns 1-6 of Fig.~\ref{fig:03} such as: (a) Ni\~no3-AISMR; (b) Ni\~no3.4-AISMR; (c) EOF1-AISMR; (d) EOF1-EOF1; (e) MCA1; (f) CCA1.} Red and blue shades correspond to $p < 0.05$, i.e., a detection of nonstationarity at that significance level.}
    \end{center}
\end{figure}

These differences may be linked to the shape and rearrangement of principal modes as described in Section~\ref{subsec:spatial}. The slight dichotomy between Figs.~\ref{fig:05}(a)-(c) and (d)-(f) may be explained by pattern rearrangements on the IM side dominantly falling outside the AISMR region but being similar for the different principal modes. Understanding the details would require better temporal resolution than in Fig.~\ref{fig:02}, but using AISMR clearly favours the detection of the long-term increase in teleconnection {\bfe strength, which is naturally explained by non-AISMR areas, especially the sea, losing importance}. On the other hand, the discrepancy between Figs.~\ref{fig:05}(d)-(e) and (f) is presumably linked to the slightly increasing importance of the middle part of the ENSO basin from the point of view of ENSO variability (reflected by EOF1, and also MCA1, as mentioned) together with the strong loss of importance of this area for the teleconnection (as CCA1 shows; see Fig.~\ref{fig:01}). It is thus not a surprise that using EOF1 or MCA1 result in lower susceptibility for detecting an increasing teleconnection strength (in comparison with Fig.~\ref{fig:05}(f)). By chance, the middle area of importance loss as seen by CCA1 coincides well with the Ni\~no3.4 and the Ni\~no3 box in the late 21st c. and around the turn of the 20th-21st centuries, respectively (see in Fig.~\ref{fig:01}), hence the respective insensitivities of these boxes to the increase and the drop in teleconnection strength in these periods (Figs.~\ref{fig:05}(b) and (a)). Note that changes in EOF1 and CCA1 are rather {\em similar} away from the middle (see again in Fig.~\ref{fig:01}), this is why the strengthening in the late 21st c. is much better caught in Fig.~\ref{fig:05}(c) than in Fig.~\ref{fig:05}(b).

Having established the observable changes, we can now ask about their drivers in terms of the linear regression model introduced in Sec.~\ref{sec:decomp}. We emphasize that we do not perform statistical tests regarding nonstationarity other than what is to do with Figs.~\ref{fig:05}, \ref{fig:06}, S8 and S9, concerning $r$ only. Therefore, the attribution of a change in $r$ to different factors is not rigorous but rather tentative.

\begin{figure}
  \begin{center}
            \includegraphics[width=\linewidth]{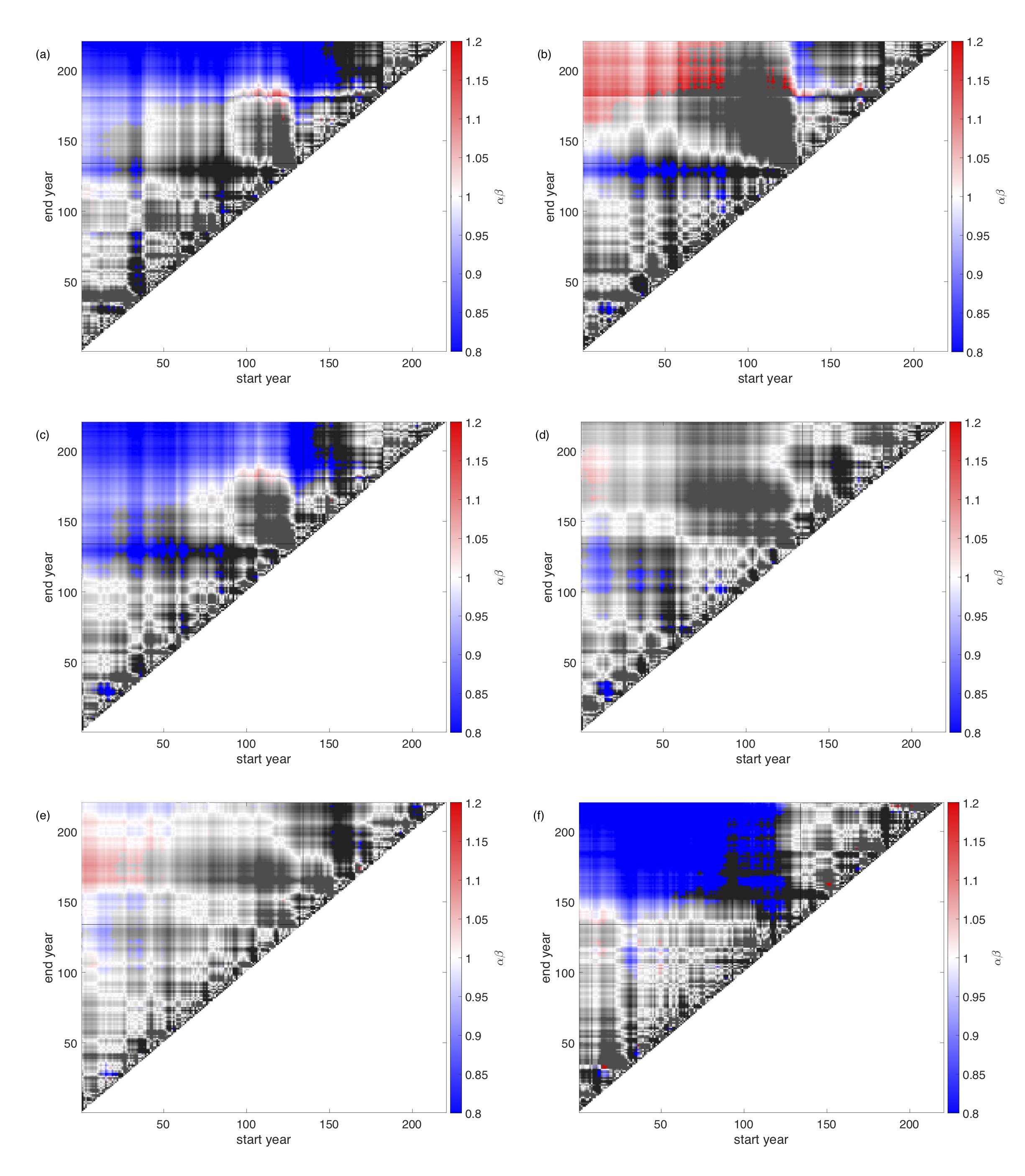} 
        \caption{\label{fig:S11} Relative importance of $\sigma_{\Phi}$ versus $a/\sigma_{\xi}$ to the change of $r(t)$. The diagrams correspond to those of Fig.~\ref{fig:05} showing the Mann-Kendall test statistics for $r(t)$. Color is applied only where $r(t)$ changes significantly. The color saturates where $\alpha\beta$ is outside of the range indicated in the colorbars.} 
    \end{center}
\end{figure}

Fig.~\ref{fig:03} suggests that ENSO variability, $\sigma_{\Phi}$, {\bfh first stagnates then increases with time in the early 21st c.} irrespective of the choice of the ENSO representation. This seems to be followed by a decrease, which, however, remains minor relative to the previous increase for Ni\~no3.4, EOF1 and MCA1. {\bfh We propose this to be the reason why} the overall increase in ENSO variability can dominate changes in the correlation coefficient by the late 21st c. for these representations of ENSO, {\bfh see in} Figs.~\ref{fig:S11}(b), (d) and (e). According to Fig.~\ref{fig:S11}, however, all other changes and all changes for other ENSO representations are more closely linked to a decrease in $a/\sigma_{\xi}$. {\bff Whether the general increase in teleconnection strength is dominated by changes in $a/\sigma_{\xi}$ or $\sigma_{\Phi}$, the 21st-c. slowdown of this increase is undoubtedly due to the decrease in $\sigma_{\Phi}$ in the same period. This must be so, since Fig.~\ref{fig:03} shows that the ratio $a/\sigma_{\xi}$ is not just decreasing for almost all representations, but there is no evident slowdown in this decrease either. {\bfh An important} nontriviality in the time evolution of $a/\sigma_{\xi}$ is} a jump at the turn of the 20th-21st centuries. This jump is related to a drop in the time evolution of $a$ and serves as a presumable explanation for the drop in the correlation coefficient at the same time. It turns out from Fig.~\ref{fig:03} that both $a$ (the inherent coupling strength) and $\sigma_{\xi}$ (the ENSO-independent part of the IM variability) typically undergo a considerable increase otherwise, with the result of $a$ ``winning'', but only slightly. The approximate balancing between these two factors may presumably be associated with both the left and right MCA1 modes being very closely resembling the corresponding EOF1 modes.\footnote{Although, the levels of correlation are noticeably different, which should be because the ``western hump'' of the MCA1 mode is somewhat more emphasized than that of the EOF1.}

{\bff Among others, we have thus seen that changes in the strength of ENSO and its coupling with the IM undergo independent changes, which may exert opposite influence on the ENSO-IM teleconnection (as represented by the Pearson correlation coefficient).} We note that there should be no apriori universal connection between an increasing ENSO variability and coupling $a$ of the ENSO and IM variability. Such a coupling to do with the precipitation in various locations can certainly have opposite trends (bottom row of Fig.~S1). %\ref{fig:S09}). 
The coupling itself should emerge as an interplay of various physical processes; both thermodynamic and dynamic processes.

\begin{figure}
%\begin{sidewaysfigure}
  %\begin{center}
  %\centering
        \includegraphics[width=1.\linewidth]{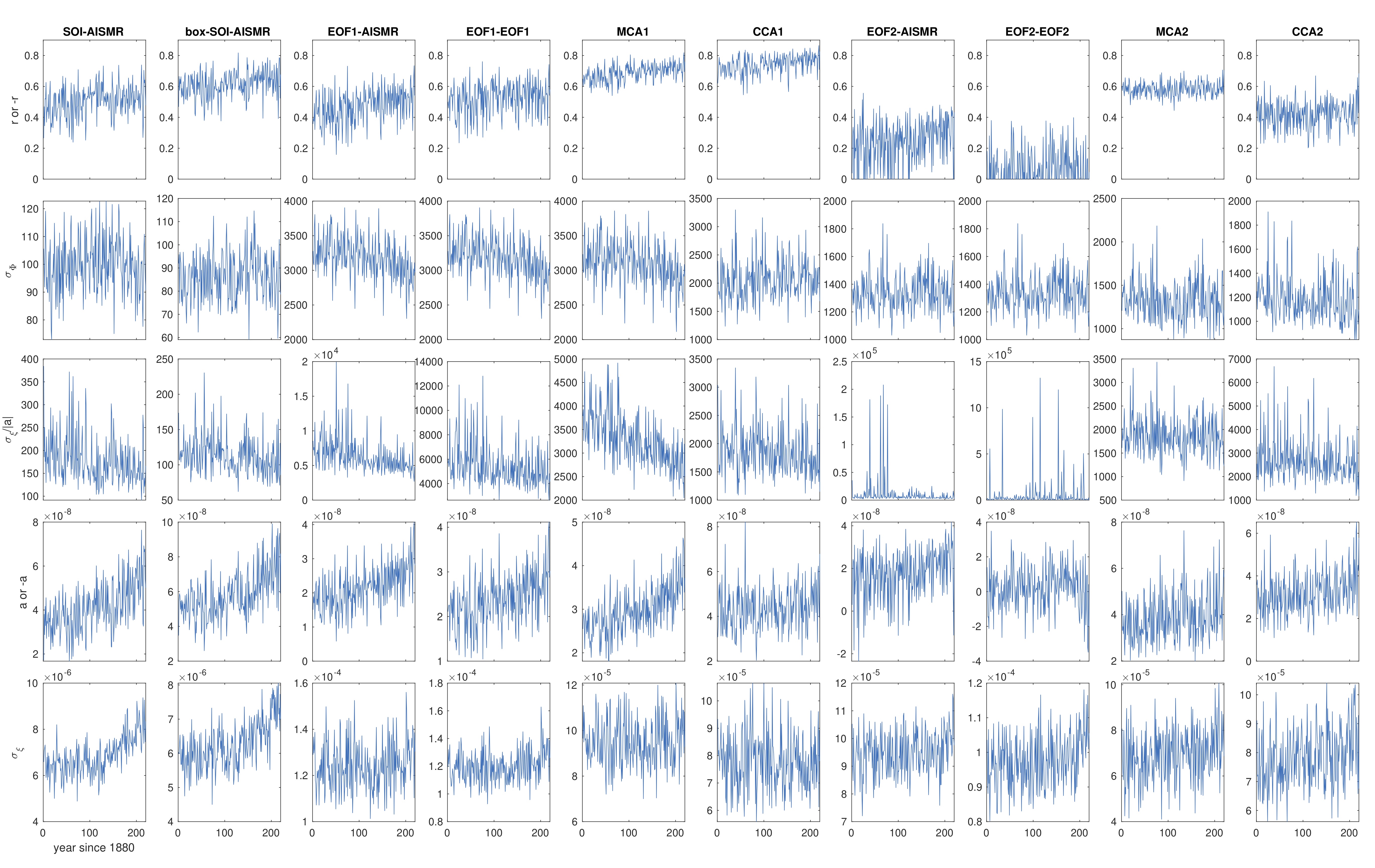}
        \caption{\label{fig:04} Same as Fig.~\ref{fig:03} but the ENSO is characterised based on the SLP. 
        }
   % \end{center}
%\end{sidewaysfigure}
\end{figure}

\begin{figure}
  \begin{center}
            \includegraphics[width=\linewidth]{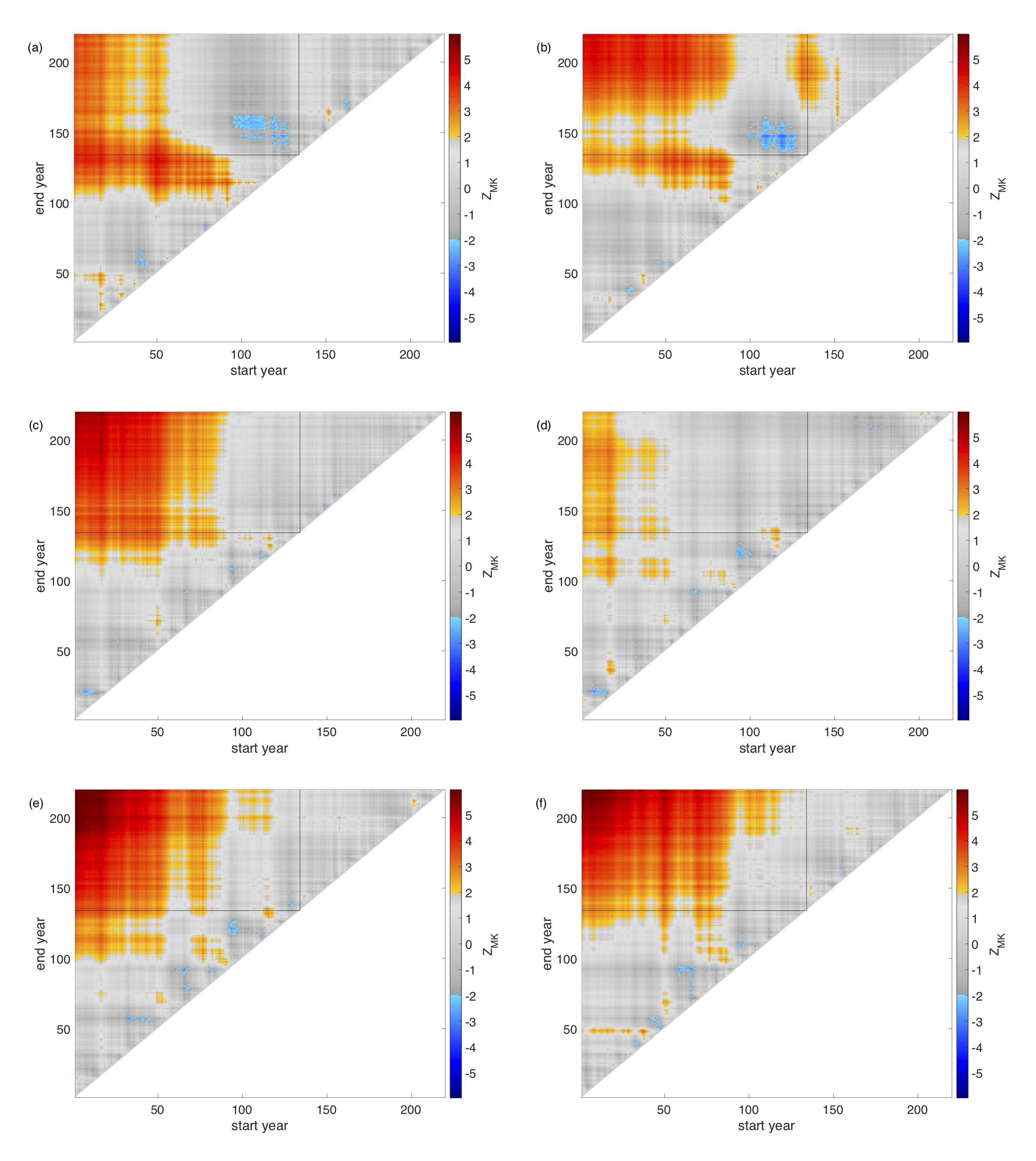} 
        \caption{\label{fig:06} Same as Fig.~\ref{fig:05} but the ENSO is characterised based on the SLP.}
    \end{center}
\end{figure}

Making use of the SLP mostly agrees with the findings obtained with the SST, see Figs.~\ref{fig:04}-\ref{fig:06}. The only but important exception is a typically decreasing ENSO variability. {\bfg We note} {\bfh that the} same domain as for SST is {\bfh obviously not suited to} SLP: the corresponding principal modes do not capture the essence of the ENSO phenomenon, as the concentration of the high-amplitude areas to the edges of the domain indicates (see Fig.~S2). %\ref{fig:S03}). 
{\bfg This should not explain the universal decrease of $\sigma_\Phi$, however, because it is decreasing also when ENSO is represented by the SOI and the box-SOI, which are defined by data outside of the used domain of the modes. Nevertheless,} we have checked (not shown) that a narrower box defining the EOF/MCA/CCA modes, extending to 10$^{\circ}$N and 10$^{\circ}$S, still leads to a {\bfg mostly decreasing} ENSO variability, with respective correlation levels largely unchanged. {\bfg Furthermore, the principal modes are not concentrated on the edges of the domain, but are largely uniform, and slightly concentrated in the middle.}

Concerning the next-to-dominant modes of variability, the correlations do not seem to feature much of a forced response (Figs.~\ref{fig:03}, \ref{fig:04}, S8, S9). %\ref{fig:S07}, S\ref{fig:S08}). 
Using the SST, the ENSO east-west dipole variability is weakening, which is the opposite of the increasing variability projected onto the dominant modes. The weakening dipole variability is more or less countered by the other factors. On the other hand, using the SLP, there seems to be no much of a change even wrt. the drivers, except for a short period at the end of the 21st c.

\section{Nonergodicity}\label{sec:nonergodicity}

Whenever the forced change is nonlinear, whether due to a nonlinear progression of the forcing or a nonlinear response characteristic, the estimation of the momentary climatology in terms of an ensemble-mean by a temporal mean in a finite window should be biased~\cite{DBT:2016}. That is, the ensemble-mean of the temporal mean would not be equal to the ensemble-mean (say, at the middle of the time window), which is termed as nonergodicity. Higher-order statistics should be biased even in the unlikely case that the forced change of the ensemble-mean is linear and the internal variability would not feature a forced change. The linear Pearson correlation coefficient is no {\bfe exception, i.e., its (traditional) time-based evaluation will be biased even if the ensemble-means of the correlated quantities exhibit a temporally linear forced response. Notwithstanding, a linear time evolution of the Pearson correlation coefficient $r(t)$ itself may lead to a vanishing bias, and this is what we will elaborate on in this section.}

Concerning the ENSO-IM teleconnection in the MPI-GE, its various {\bfe representations, in fact, clearly} feature a nonlinear change of $r(t)$, whether it is a monotonic but degressive change (represented by a concave graph) or a nonmonotonic one, as seen in Figs.~\ref{fig:03}, \ref{fig:04}, \ref{fig:05}, \ref{fig:06}. However, the evaluation of the bias or the degree of nonergodicity is not straightforward in this case since the ``signal-to-noise ratio'' is rather poor thanks to the relatively small ensemble size. Nevertheless, we can attempt to at least {\em detect} nonergodicity by performing a statistical test. The quantity of the so-called test statistics can in turn serve as some {\em quantifier} of the degree of nonergodicity. 

To the end of constructing a suitable test, we observe that the Fisher transform $\hat{z}=\arctanh(\hat{r})$ (e.g. $\hat{r}$ distinguishes the finite-$N$ sample estimate of $r$) provides approximately normally distributed iid rv's (wrt. the different years) of standard deviation approximately $1/\sqrt{N-3}$ for large enough $N$ and {\em any} true value $r$ of the correlation coefficient.\footnote{\bfe This approximation assumes the correlated variables to follow a Gaussian distribution, but non-Gaussianity has been checked to have a farly negligible effect for some basic ENSO-quantifiers and AISMR in~\cite{BDHLL:2020}.} The same applies, of course, to the finite-$\tau$-window sample correlation coefficient $r_{\tau}$ when $\Phi(t)$, $\Psi(t)$ are not {\bfe auto-correlated and their ensemble-means and standard deviations remain constant in time}.  
{\bfe In such a case}, the only reason that the expectation 
\begin{equation}\label{eq:std_est_inst}
   \langle(\hat{z}_{\tau}-z)^2\rangle
\end{equation}
at any time does not match the theoretical value $(\tau-3)^{-1}$ is the existence of a bias:
\begin{equation}
   \langle \hat{z}_{\tau}\rangle\neq z,	
\end{equation}
i.e., nonergodicity. We have {\bfa four} issues to consider. First, $z$ is not available because of the finite $N$ ensemble size. Therefore, we cannot evaluate nonergodicity at any time (every year). Instead, we can concern nonergodicity overall in an interval of length $T$ calculating 
\begin{equation}\label{eq:std_est}
   v = \sum_{t=1}^T\sum_{n=1}^N(\hat{z}_{\tau,t,n}-\hat{z}_t)^2.
\end{equation}
Second, had $\hat{z}_{\tau}$ and $\hat{z}$ in the above been {\bfe computed from} combinations (without repetition) from the same sample of size $T=N$, Cochran's theorem~\cite{cochran_1934} would dictate that $cv$ (where $c$ is an appropriate constant factor) would be distributed according to a $\chi^2$-distribution. This would allow for calculating the p-value the usual way by evaluating the $\chi^2$-distribution at the level given by the calculated test statistics. However, our setting is somewhat different, which might not be possible to tackle analytically. Nevertheless, one can sample the test statistics and, therefore, determine its quantiles to arbitrary precision. We have done this sampling simply by generating sample correlation realisations by generating realisations of correlated random variables $X$ and $Y$, where $Y = aX + \xi$, and $X$ and $\xi$ are normally distributed independent rv's. {\bfe This is a further assumption for the nature of our variables, from which deviations certainly exist but which are presumably moderate enough to have a secondary effect for the results.} Note that the moving-window temporal correlation coefficient $\hat{r}_{\tau,t}$ is calculated upon pregenerating $x_t$ and $y_t$, $t=1,\dots,T$. We have checked that $cv$ does seem to follow a $\chi^2$-distribution even in our setting. Third, ENSO is a dynamical phenomenon and $\Phi$, in any representation, is in fact auto-correlated. We have checked that taking $X$ to be governed by an auto-regressive process of order 1, $X_{t+1}=\phi X_t+\xi_x$, such that the lag-1 autocorrelation is 0.3, shifts the distribution of $v$ little wrt. its std. We use this value of 0.3 in order to calculate the p-value, or, the 0.95 quantile of $v$. The latter is about 670 using the parameters $T=220$, $\tau=31$, $N=63$; {\bfa while having a serially uncorrelated $X_t$, it is 667. As a fourth issue, there might be a low-frequency influence on the teleconnection, say, via a time-dependent $a$~\cite{Gershunov:1998,TW:1999,KG:2000,Krishnamurthy:2014,Watanabe:2014}. This should increase the width of the distribution of the sample temporal correlation coefficient, i.e., it should be larger than $1/\sqrt{N-3}$. We tested the effect of this by introducing an additive perturbation, $a_t=a_0 + \delta a_t$, where $\delta a_t$ is modeled again as an AR(1) process, setting $\phi=0.8$ and {\bfe such} a noise {\bfe strength that} std$[\delta a_t]=0.05$. With this the 0.95 quantile is 674; that is, the effect is rather small. This corroborates well with the findings of~\cite{Gershunov:2001,YT:2018}, namely, that even if there was a low-frequency modulation of the ENSO-IM teleconnection strength, it is too weak to detect from century-long observations. The cancellation effect described by~\cite{KG:2000} may be an alternative view of this.}

\begin{figure}
  \begin{center}
        \includegraphics[width=\linewidth]{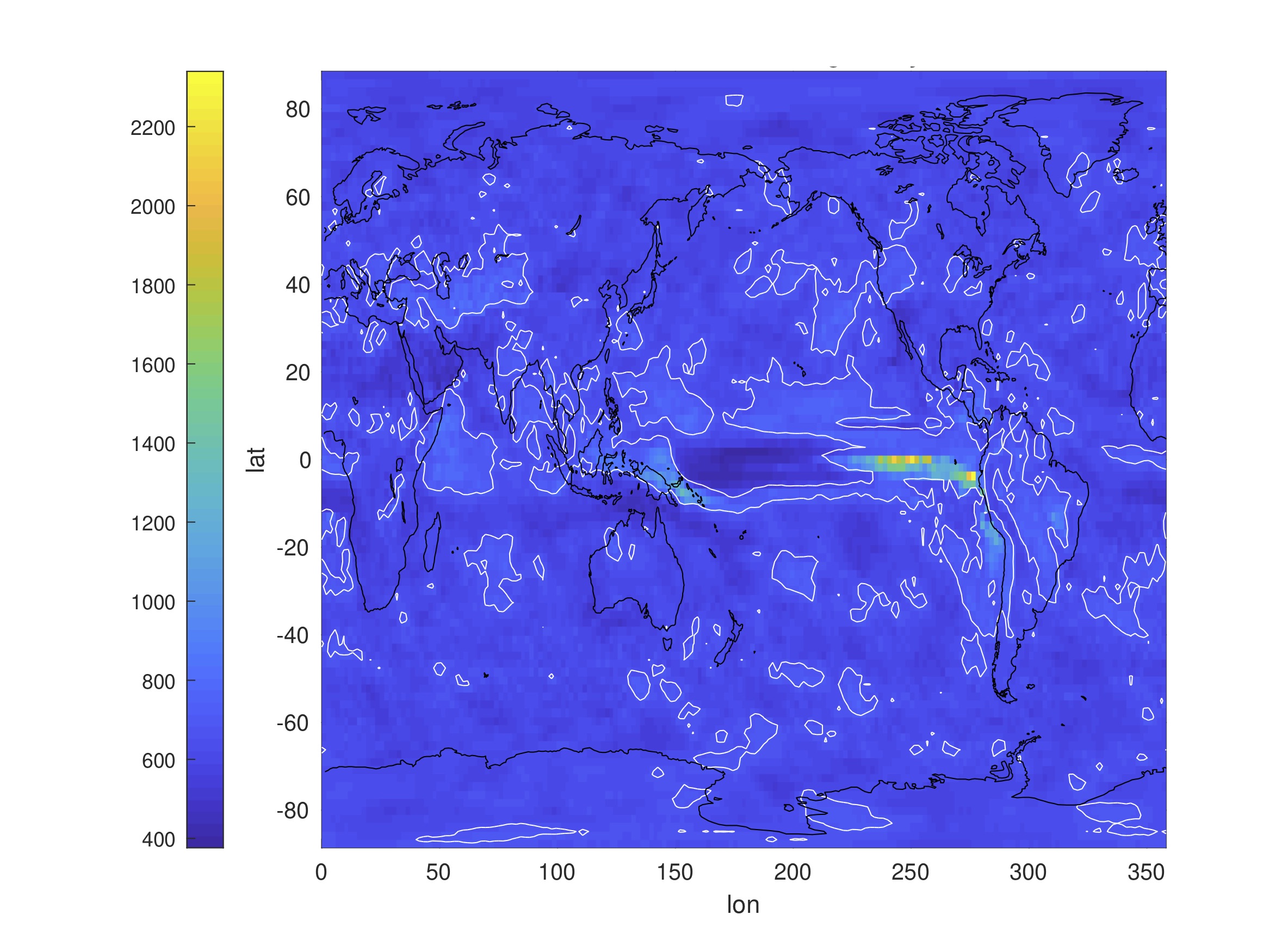}
        \caption{\label{fig:07} Test statistics of our ad hoc test of nonergodicity regarding the Pearson correlation coefficient. The contour marks the 0.95 quantile.}
    \end{center}
\end{figure}

Turning to our application, the test statistics is found to be 756 for the ENSO-IM teleconnection representation given by the Ni\~no3-ISMR pair. That is, it is well above the 0.95 quantile, corresponding to a minuscule p-value, so that the hypothesis of ergodicity can be rejected with extremely high confidence. We also evaluate the test statistics corresponding to the global correlation maps seen in Fig.~S1. %\ref{fig:S09}. 
The result of this can be presented as a global map too (Fig.~\ref{fig:07}), in which a contour for the 0.95 quantile 670 encloses regions where ergodicity can be rejected at the usual significance level. We can see such regions not just in the tropics or in the Equatorial Pacific, but all over the world. Nevertheless, the highest levels of bias/nonergodicity is found indeed at the centre of ENSO variability and elsewhere in the Equatorial Pacific. 

We note that even if $r(t)$ changes linearly, or not at all, there could be a trivial source of the bias, namely, a changing ensemble-mean or standard deviation of $\Phi$ or $\Psi$. We redo the calculation of the test statistics in order to eliminate this trivial source. It is done by subtracting the respective ensemble-mean time series from the $\Phi$ and $\Psi$ time series, separately for each ensemble member, before calculating the temporal correlation coefficient $\hat{r}_{\tau}$. The result (Fig. S10) %\ref{fig:S10}) 
is hardly distinguishable from the original one (Fig.~\ref{fig:07}) with respect to the patterns, only the values of $v$ are slightly off. That is, in this case the change of the mean state hardly {\bfe contributes} to the bias. {\bfg In fact, the forced change of the standard deviation can also be a source of bias. We believe that just as the ensemble-mean change, this is also a negligible effect, considering especially that in $\tau=30$ yr windows the changes of the $\sigma_{\Phi}$'s in Figs.~\ref{fig:03} and \ref{fig:04} seem unlike to be detectable.} For this reason, nonergodicity {\bfg should robustly} imply the nonlinearity of $r(t)$.

% \begin{figure}
%   \begin{center}
%         \includegraphics[width=\linewidth]{ENSO_IM_teleco_17_sqmap}
%         \caption{\label{fig:07} Test statistics of our ad hoc test of nonergodicity regarding the Pearson correlation coefficient. The contour marks the 0.95 quantile.}
%     \end{center}
% \end{figure}

\section{Discussion and conclusions}\label{sec:discussion}

We have re-examined the forced response of the ENSO-Indian monsoon (IM) teleconnection as conveyed by the MPI-GE data. One main increment taken by the new analysis is the consideration of spatial aspects of any forced change. This was achieved by determining the dominant as well as the next-to-dominant modes of variability in terms of two leading EOFs and modes of Maximum Covariance Analysis (MCA) and Canonical Correlation Analysis (CCA). In turn, the current teleconnection strength was defined by the ensemble-wise Pearson correlation coefficient between the scalar data series obtained by projecting SST/SLP and precipitation fields {\bff onto} any of the modes. The plurality of the representation of the ENSO-IM teleconnection allowed us {\bff to build} a robust picture of forced change.

We have found that the dominant modes of variability in all {\bff representations, similarly to the classical ENSO indices,} convey a picture of a strengthening teleconnection in terms of a statistical test. {\bff However, a slight drop around the turn of the 20th-21st centuries also turns out to be present irrespective of the particular representation, and a late-21st-c. slowdown in the increase of the strength as well.}
% However, in the 21st c. under the strong RCP8.5 forcing scenario there is typically less change than in the historical period. Some of this
{\bff The latter mostly has} to do with a curious nonmonotonicity in the change of the ENSO variability in most representations: about midway in the 21st c. rather suddenly it starts to decline. In terms of a linear regression model that can be associated with evaluating the (linear) Pearson correlation coefficient, the typically increasing regression coefficient -- which can be viewed as an ENSO-IM coupling strength -- {\bff also plays a strong role. Although the model's noise strength undergoes a similar increase, which has an opposing role, the change of the coupling is found to dominate.} The latter turns out to be the central piece of the robust or consistent picture of the strengthening ENSO-IM teleconnection. {\bff We conjecture that a temporary drop in the coupling strength at the turn of the 20th-21st centuries also has an important effect: a slight decrease in the Pearson correlation coefficient at that time.}

We leave it for future work to attribute {\bff these statistical features to physical effects.} However, the change of both the coupling and noise strength should involve both thermodynamic and dynamic factors. {\bff We emphasize that all these findings are identified rather robustly across several representations of the ENSO-IM teleconnection. We must also admit, however, that some of these features are not pronounced in all representations and may even not pass a strict detectability threshold in some of them.}

{\bff These differences between different representations are due to changes in the relevant spatial patterns. The most interesting finding is that the pattern of maximal ENSO variability and that of maximal correlation with the IM precipitation undergo opposing trends in the middle of the ENSO domain: the middle becomes more important for ENSO itself, while it loses importance for the teleconnection. On the side of the IM, however, the intrinsic and teleconnective patterns change very similarly. In particular, most areas over the sea lose much of their importance for both aspects.}

It is {\bff essential} to note that based on observations the ENSO-IM teleconnection appears to have weakened since about 1980~\cite{KK:1999,BDHLL:2020}. In~\cite{BDHLL:2020} we argued that several reasons could be responsible for this mismatch. Among them one is that the MPI-ESM associated with the MPI-GE is not consistent with the observations of the 20th c. Indian Ocean (IO) warming and IM precipitation decline: in terms of the long-term trend, the ranges of simulated realisations, wrt. both the IO temperature and ISMR, do not contain the observations as respective single realisations -- by far (see Figs. 11 and 13 of~\cite{BDHLL:2020}). See~\cite{Aneesh:2018}, for example, for the role of the IO warming regarding the decline of ISMR, concerning La Ni\~na years, in particular. However, it is a question yet to be answered if the decline of precipitation does actually translate into a weakening of the ENSO-IM teleconnection in the precise sense of a {\em forced} change.

Another possibility for the reason for the said mismatch was posited in~\cite{BDHLL:2020} to be the very considerable fluctuations of temporal correlation coefficients -- the larger in magnitude the shorter the time window -- even under stationary unforced conditions, and even without decadal variability of the correlated signals. Evaluating temporal correlations is troubled by further problems when there is a forced change, namely, that (i) when the forced change is nonlinear, the temporal correlations are biased (Sec.~\ref{sec:nonergodicity}), and that (ii) the forced signal is not known, and, therefore, the anomalies that are meant to be correlated cannot even be constructed. A detrending, say, removing a linear slope in a time window, does not seem to be a good fix, as we do not know whether there is any forced change at all, being the very question that we are asking, and even under stationary conditions one would find spurious trends due merely to internal variability. 

When an ensemble is available concerning a {\bff single} model at least, the biases to do with point (i) above can be detected and, to a certain extent, quantified. We have developed an ad hoc statistical test to do just that, and applied it to the ENSO-IM teleconnection as well as the relationship of global precipitation and ENSO. We have found many regions of the world where a bias could be detected. The reasons for the nonlinearities that such biases imply can be manifold, not only driven by the nonlinearity of the change of ENSO variability, and it should be carefully examined in the future. 

{\bff Our methodology indeed ensures that the detected biases imply nonlinearity (i.e., not only nonlinearity implies biases, but also the other way round). In our context of the historical and scenario change of ENSO-precipitation teleconnections, robust nonlinearity has been found.} Given that nonlinearity implies (presupposes) a forced change, we turn out to have on hand a statistical test for the nonstationarity of teleconnections (linear correlations). This can be seen to complement the Mann-Kendall test (in the special case of the nonstationarity of the correlation coefficient), because the latter takes the alternative hypothesis of a monotonic change, while our test includes nonmonotonic changes too.

\section*{Conflict of Interest Statement}
%All financial, commercial or other relationships that might be perceived by the academic community as representing a potential conflict of interest must be disclosed. If no such relationship exists, authors will be asked to confirm the following statement: 

The authors declare that the research was conducted in the absence of any commercial or financial relationships that could be construed as a potential conflict of interest.

\section*{Author Contributions}

%The Author Contributions section is mandatory for all articles, including articles by sole authors. If an appropriate statement is not provided on submission, a standard one will be inserted during the production process. The Author Contributions statement must describe the contributions of individual authors referred to by their initials and, in doing so, all authors agree to be accountable for the content of the work. Please see  \href{http://home.frontiersin.org/about/author-guidelines#AuthorandContributors}{here} for full authorship criteria.

TB developed any new code, including those implementing SMCA and SCCA, carried out all the numerical analyses and created all the figures for the paper. GD and subsequently TB originated the ideas of SMCA and SCCA, respectively. TB proposed the breaking down of the change of the correlation coefficients into three constituents according to the linear regression model. TB devised the statistical test for nonergodicity of correlations motivated by Axel Timmermann's enquiry. GD, J-Y L, E-S C suggested to calculate the differences as e.g. in Fig.~\ref{fig:01}; and to evaluate the relative importance of ENSO variability as seen in Fig.~\ref{fig:S11} developed and implemented by TB. TB and GD lead the writing of the manuscript and the interpretation of results. K-J H, J-Y L, E-S C contributed to the manuscript. 

\section*{Funding}
%Details of all funding sources should be provided, including grant numbers if applicable. Please ensure to add all necessary funding information, as after publication this is no longer possible.

TB was supported by the Institute for Basic Science (IBS), under IBS-R028-D1. 
G.D. acknowledges financial support from the Spanish State Research Agency through the Mar\'{\i}a de Maeztu Program for Units of Excellence in R\&D under grant no. MDM-2017-0711, from the National Research, Development and Innovation Office (NKFIH, Hungary) under grant no. K125171, and from the European Social Fund through the fellowship no. PD/020/2018 under the postdoctoral program of the Government of the Balearic Islands (CAIB, Spain).

\section*{Acknowledgments}
%This is a short text to acknowledge the contributions of specific colleagues, institutions, or agencies that aided the efforts of the authors.

TB thanks Malte Stuecker for useful discussions and Axel Timmermann for 1. his interest in nonergodicity and 2. discussing whether nonergodicity implies the nonlinear change of correlations and vice versa. TB thanks Keith Rodgers for informing him about the volume of {\em Variations} that includes~\cite{Timi_Variations}. GD is thankful to B. Stevens, T. Mauritsen, Y. Takano, and N. Maher for providing access to the output of the MPI-ESM ensembles.

\section*{Supplemental Data}
 %\href{http://home.frontiersin.org/about/author-guidelines#SupplementaryMaterial}{Supplementary Material} should be uploaded separately on submission, if there are Supplementary Figures, please include the caption in the same file as the figure. LaTeX Supplementary Material templates can be found in the Frontiers LaTeX folder.
 
Supplementary Figures are available online, on the webpage of the article. Supplementary Videos are available online on YouTube, as cited in the text. % not from the article webpage because they only allow 30MB.

\section*{Data Availability Statement}
%The datasets [GENERATED/ANALYZED] for this study can be found in the [NAME OF REPOSITORY] [LINK].
% Please see the availability of data guidelines for more information, at https://www.frontiersin.org/about/author-guidelines#AvailabilityofData

The data generated and analysed for this study are available upon reasonable request. For reproducing purposes, one can make use of the data available at \url{https://esgf-data.dkrz.de/projects/mpi-ge/}, as suggested in Sec.~\ref{sec:methods}.

\bibliographystyle{unsrt}
%\bibliography{ENSO_telecons_nonergod}

%%% Make sure to upload the bib file along with the tex file and PDF
%%% Please see the test.bib file for some examples of references

\newpage

%\section*{Figure captions}

%%% Please be aware that for original research articles we only permit a combined number of 15 figures and tables, one figure with multiple subfigures will count as only one figure.
%%% Use this if adding the figures directly in the mansucript, if so, please remember to also upload the files when submitting your article
%%% There is no need for adding the file termination, as long as you indicate where the file is saved. In the examples below the files (logo1.eps and logos.eps) are in the Frontiers LaTeX folder
%%% If using *.tif files convert them to .jpg or .png
%%%  NB logo1.eps is required in the path in order to correctly compile front page header %%%

% \begin{figure}[h!]
% \begin{center}
% \includegraphics[width=10cm]{logo1}% This is a *.eps file
% \end{center}
% \caption{ Enter the caption for your figure here.  Repeat as  necessary for each of your figures}\label{fig:1}
% \end{figure}
% 
% 
% \begin{figure}[h!]
% \begin{center}
% \includegraphics[width=15cm]{logos}
% \end{center}
% \caption{This is a figure with sub figures, \textbf{(A)} is one logo, \textbf{(B)} is a different logo.}\label{fig:2}
% \end{figure}

%%% If you are submitting a figure with subfigures please combine these into one image file with part labels integrated.
%%% If you don't add the figures in the LaTeX files, please upload them when submitting the article.
%%% Frontiers will add the figures at the end of the provisional pdf automatically
%%% The use of LaTeX coding to draw Diagrams/Figures/Structures should be avoided. They should be external callouts including graphics.

%\begin{sidewaysfigure}
%\begin{landscape}
%\end{sidewaysfigure}

% \begin{figure}
%   \begin{center}
%         \begin{tabular}{cc}
%             \includegraphics[width=0.5\linewidth]{ENSO_IM_teleco_12_contrib_ZMK_r_sca_e.jpg} 
%             \includegraphics[width=0.5\linewidth]{ENSO_IM_teleco_12_contrib_ZMK_r_sca_e_34.jpg} \\
%             \includegraphics[width=0.5\linewidth]{ENSO_IM_teleco_12_contrib_ZMK_r_eof1_e.jpg} 
%             \includegraphics[width=0.5\linewidth]{ENSO_IM_teleco_12_contrib_ZMK_r_eof11_e.jpg}
%             \\
%             \includegraphics[width=0.5\linewidth]{ENSO_IM_teleco_12_contrib_ZMK_r_mca.jpg} 
%             \includegraphics[width=0.5\linewidth]{ENSO_IM_teleco_12_contrib_ZMK_r_cca.jpg}
%         \end{tabular}
%         \caption{\label{fig:05} Test statistics of the Mann-Kendall test for the stationarity of $r(t)$ {\bfa or $-r(t)$, depending on which of these gives an overall positive value}. {\bfg The diagrams correspond to columns 1-6 of Fig.~\ref{fig:03} such as: (a) Ni\~no3-AISMR; (b) Ni\~no3.4-AISMR; (c) EOF1-AISMR; (d) EOF1-EOF1; (e) MCA1; (f) CCA1.} Red and blue shades correspond to $p < 0.05$, i.e., a detection of nonstationarity at that significance level.}
%     \end{center}
% \end{figure}

% 25.06.20.
% \begin{figure}
%   \begin{center}
%         \begin{tabular}{cc}
%             \includegraphics[width=0.5\linewidth]{ENSO_IM_teleco_12_contrib_rto_r_sca_e.jpg} 
%             \includegraphics[width=0.5\linewidth]{ENSO_IM_teleco_12_contrib_rto_r_sca_e_34.jpg} \\
%             \includegraphics[width=0.5\linewidth]{ENSO_IM_teleco_12_contrib_rto_r_eof1_e.jpg} 
%             \includegraphics[width=0.5\linewidth]{ENSO_IM_teleco_12_contrib_rto_r_eof11_e.jpg} \\
%             \includegraphics[width=0.5\linewidth]{ENSO_IM_teleco_12_contrib_rto_r_mca.jpg} 
%             \includegraphics[width=0.5\linewidth]{ENSO_IM_teleco_12_contrib_rto_r_cca.jpg} 
%         \end{tabular}
%         \caption{\label{fig:S11} Relative importance of $\sigma_{\Phi}$ versus $a/\sigma_{\xi}$ to the change of $r(t)$. The diagrams correspond to those of Fig.~\ref{fig:05} showing the Mann-Kendall test statistics for $r(t)$. Color is applied only where $r(t)$ changes significantly. The color saturates where $\alpha\beta$ is outside of the range indicated in the colorbars.} 
%     \end{center}
% \end{figure}

% 1308.2020. Latex will not see the edit of ENSO_IM_teleco_12_decomp_new-eps-converted-to.pdf by preview. I tried to put a text box over the figure title to cover out the old one. Have rerun ENSO_IM_teleco_12_decomp.m and ENSO_IM_teleco_14_decomp.m on Alpeh, after correcting the scripts reg' the fig' titles.

% \begin{figure}
%   \begin{center}
%         \begin{tabular}{cc}
%             \includegraphics[width=0.5\linewidth]{ENSO_IM_teleco_14_contrib_ZMK_r_sca_e.jpg} 
%             \includegraphics[width=0.5\linewidth]{ENSO_IM_teleco_14_contrib_ZMK_r_sca_e_bx.jpg} \\
%             \includegraphics[width=0.5\linewidth]{ENSO_IM_teleco_14_contrib_ZMK_r_eof1_e.jpg} 
%             \includegraphics[width=0.5\linewidth]{ENSO_IM_teleco_14_contrib_ZMK_r_eof11_e.jpg}
%             \\
%             \includegraphics[width=0.5\linewidth]{ENSO_IM_teleco_14_contrib_ZMK_r_mca.jpg} 
%             \includegraphics[width=0.5\linewidth]{ENSO_IM_teleco_14_contrib_ZMK_r_cca.jpg}
%         \end{tabular}
%         \caption{\label{fig:06} Same as Fig.~\ref{fig:05} but the ENSO is characterised based on the SLP.}
%     \end{center}
% \end{figure}

\newpage

\renewcommand{\figurename}{Figure S}
\setcounter{figure}{0}

\section*{Supplementary Material}

\begin{figure}
  \begin{center}
            \includegraphics[width=\linewidth]{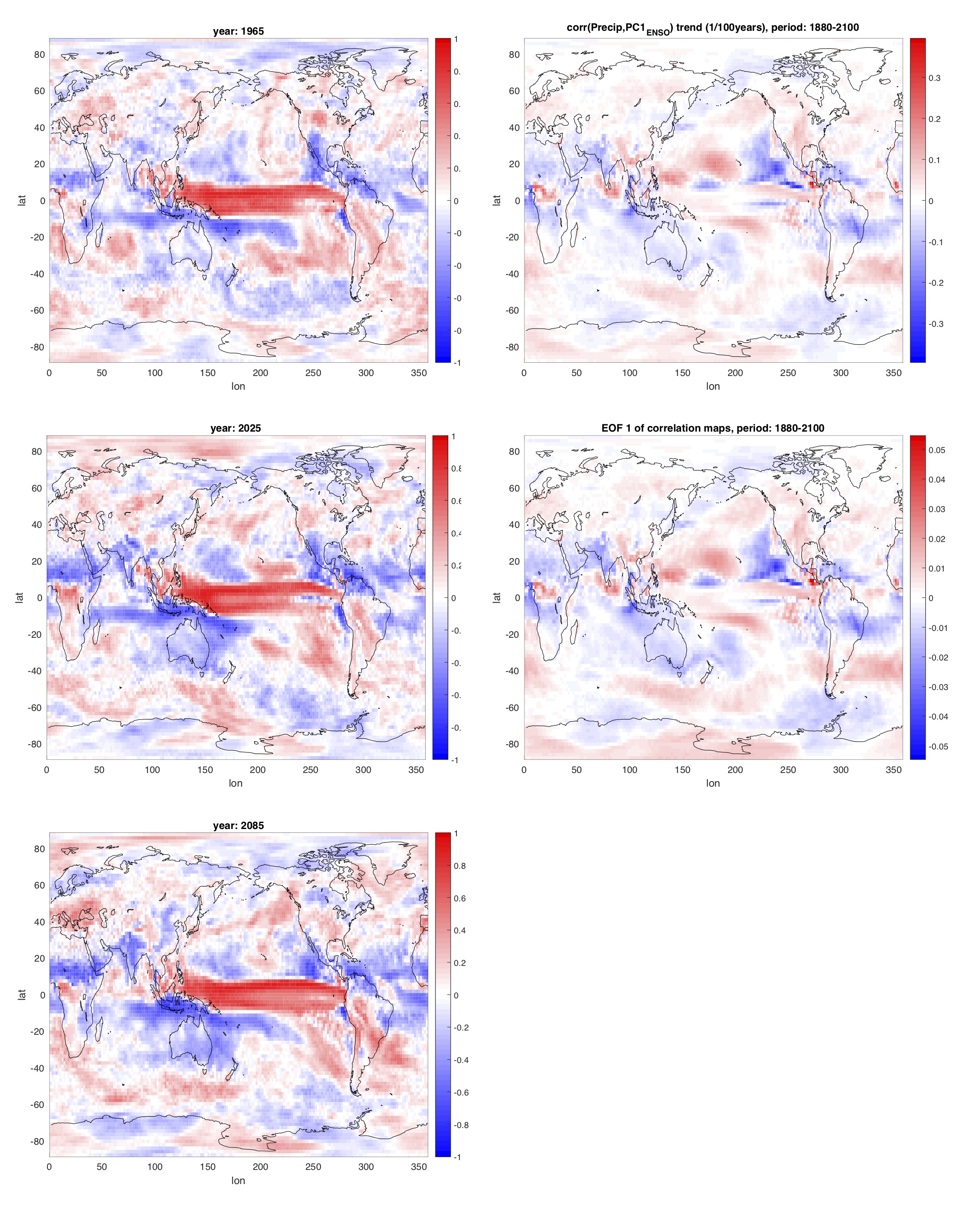}
        \caption{\label{fig:S09} Sample snapshot correlation maps at the indicated years (left column; the same is shown in Fig. 4a-c of~\cite{HHB:2020} concerning the CESM1-LE) and corresponding linear trends (top right; the same is shown in Fig. 5a of~\cite{HHB:2020}) and temporal EOF1 for the full evolution of the correlation maps (middle right). We note that if the response (characteristic) is linear in every gridpoint and the forcing is quasi-statically slow, then the single eigenvalue that is nonzero is the one belonging to EOF1; the forced response of the field is a perfect standing wave. JJA-mean SST and precipitation data is used.
        }
    \end{center}
\end{figure}

\begin{figure*}
  \begin{center}
      \includegraphics[width=\linewidth]{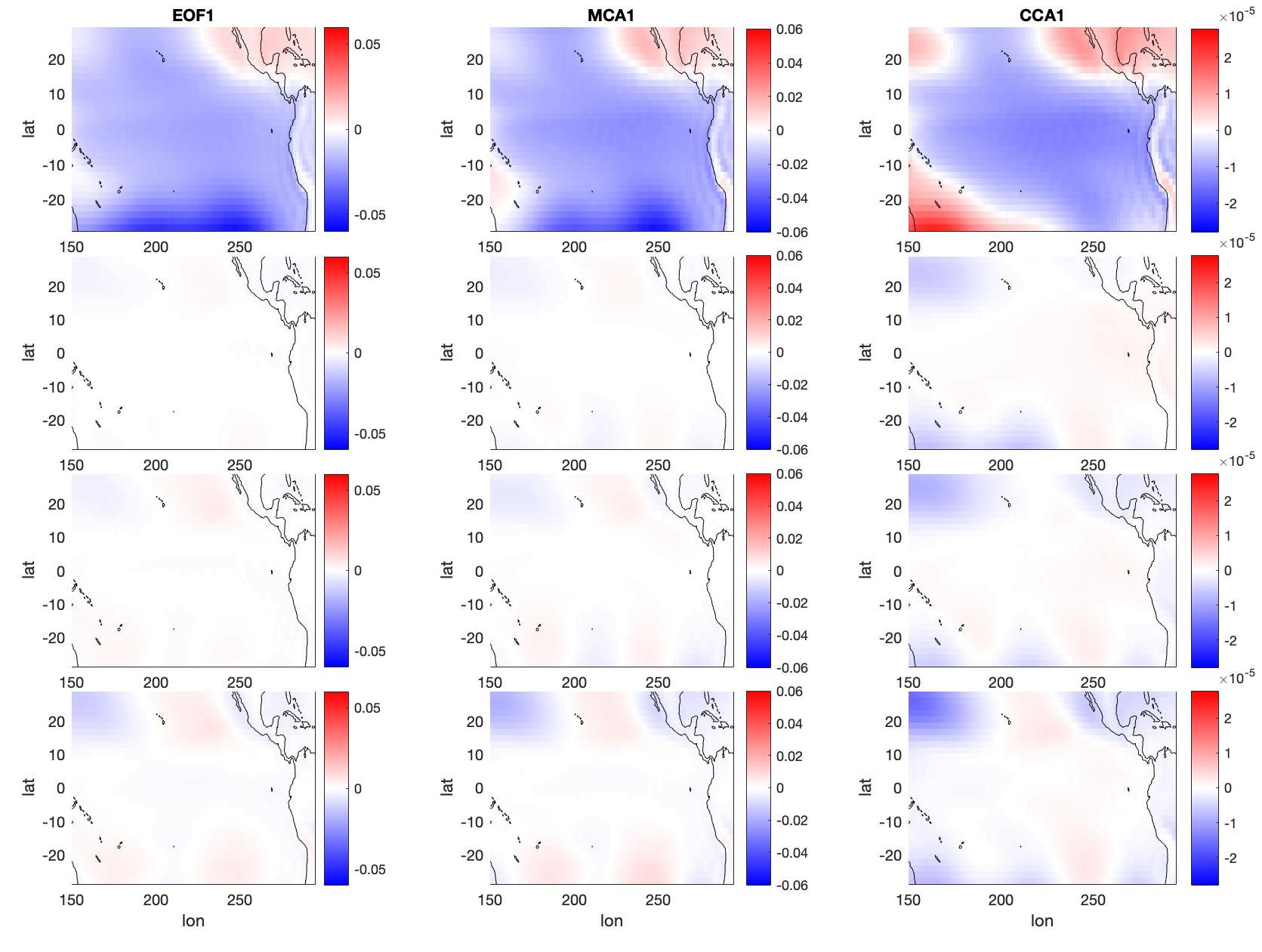} 
      \caption{\label{fig:S03} Same as Fig. 1 %~\ref{fig:01} 
      (dominant modes) but the ENSO is characterised by the SLP.}
  \end{center}
\end{figure*}

\begin{figure*}
  \begin{center}
      \includegraphics[width=\linewidth]{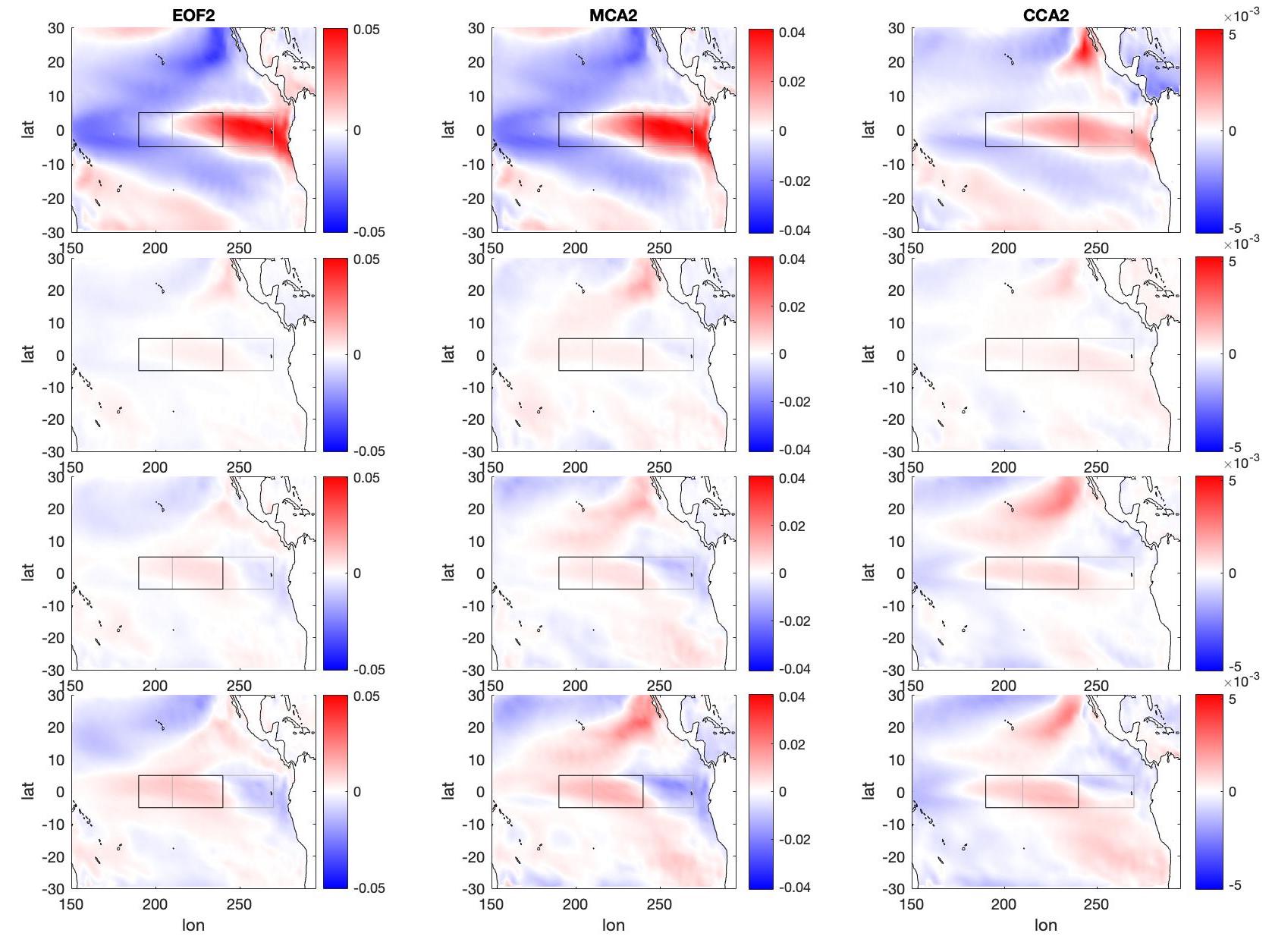} 
      \caption{\label{fig:S01} Same as Fig. 1 %~\ref{fig:01} 
      (SST) but for the next-to-dominant modes.}
  \end{center}
\end{figure*}

\begin{figure*}
  \begin{center}
      \includegraphics[width=1.\linewidth]{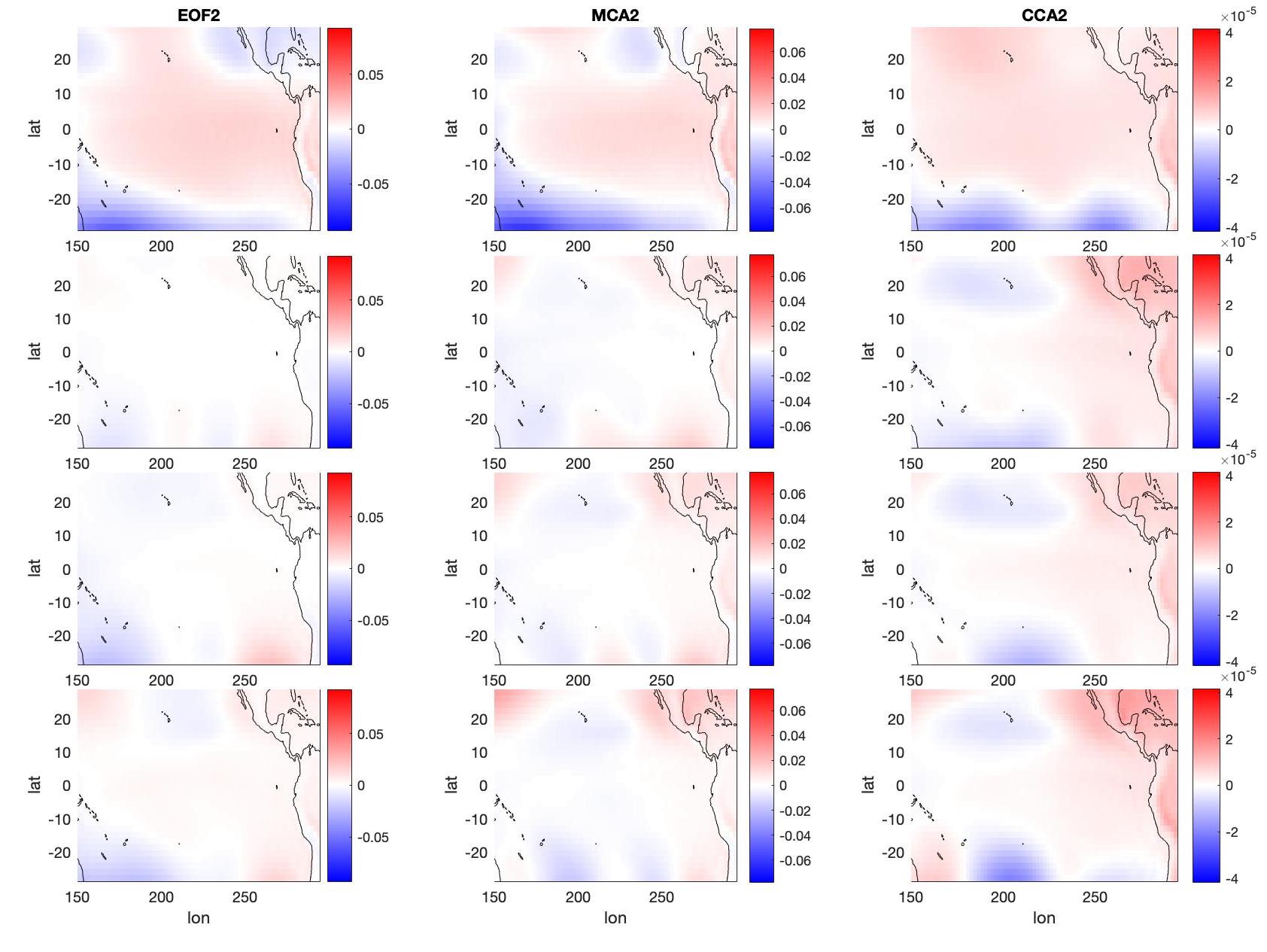} 
      \caption{\label{fig:S04} Same as Fig.~\ref{fig:S03} (SLP) but for the next-to-dominant modes.}
  \end{center}
\end{figure*}

\begin{figure*}
  \begin{center}
      \includegraphics[width=\linewidth]{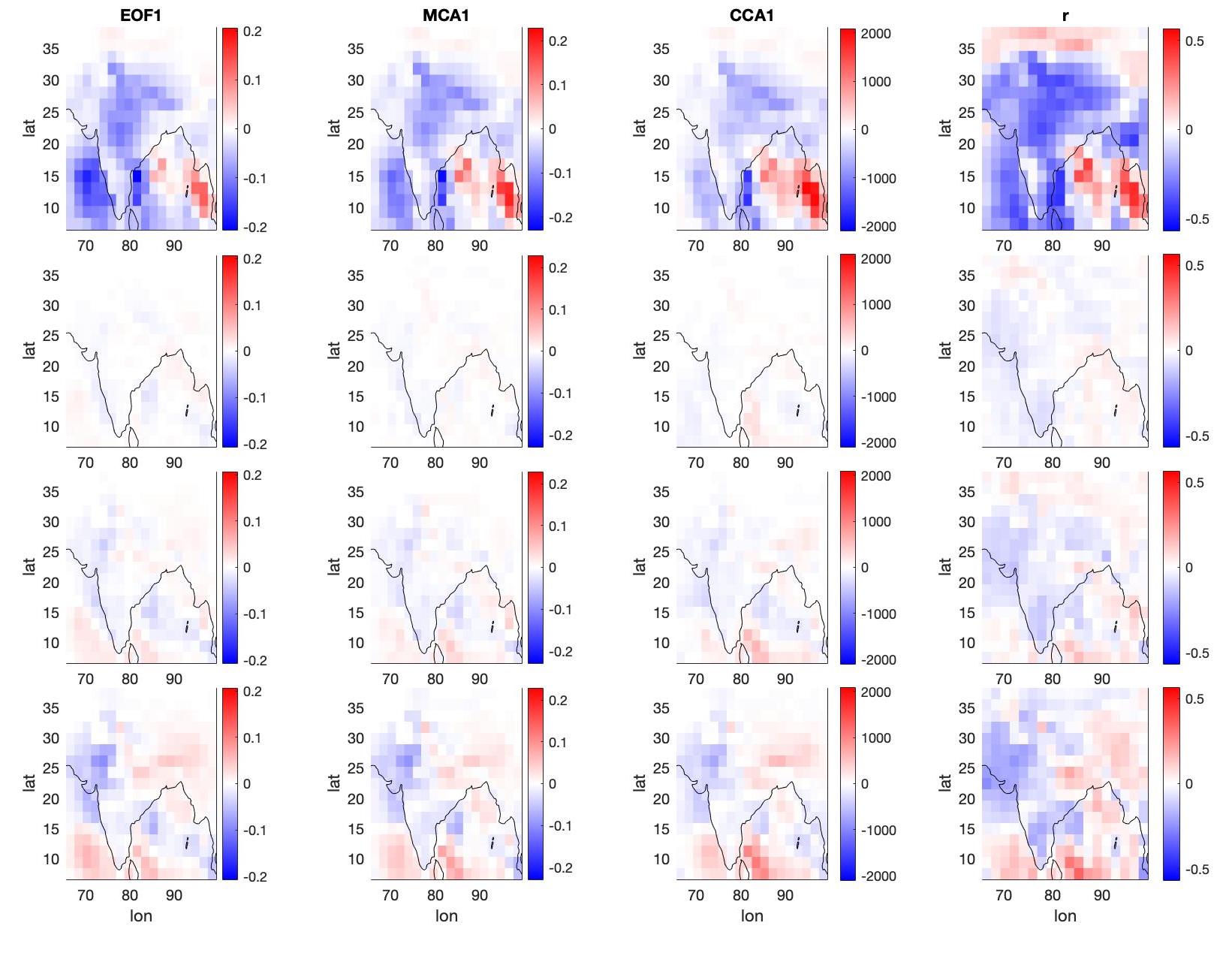} 
      \caption{\label{fig:S05} Same as Fig. 2 %~\ref{fig:02} 
      (dominant modes) but the ENSO is characterised by the SLP.}
  \end{center}
\end{figure*}

\begin{figure*}
  \begin{center}
      \includegraphics[width=1.\linewidth]{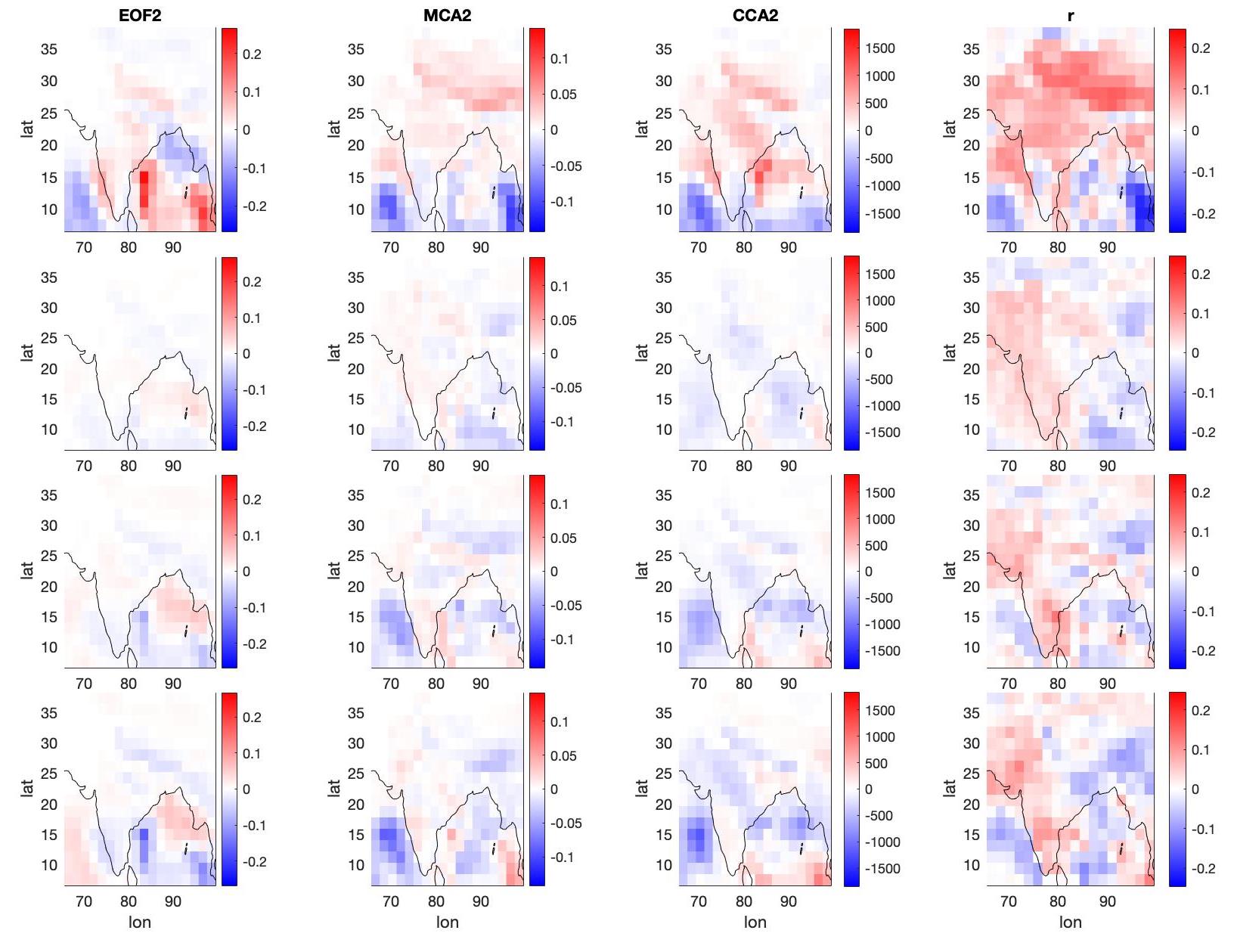} 
      \caption{\label{fig:S02} Same as Fig. 2 %~\ref{fig:02} 
      but for the next-to-dominant modes.}
  \end{center}
\end{figure*}

\begin{figure*}
  \begin{center}
      \includegraphics[width=1.\linewidth]{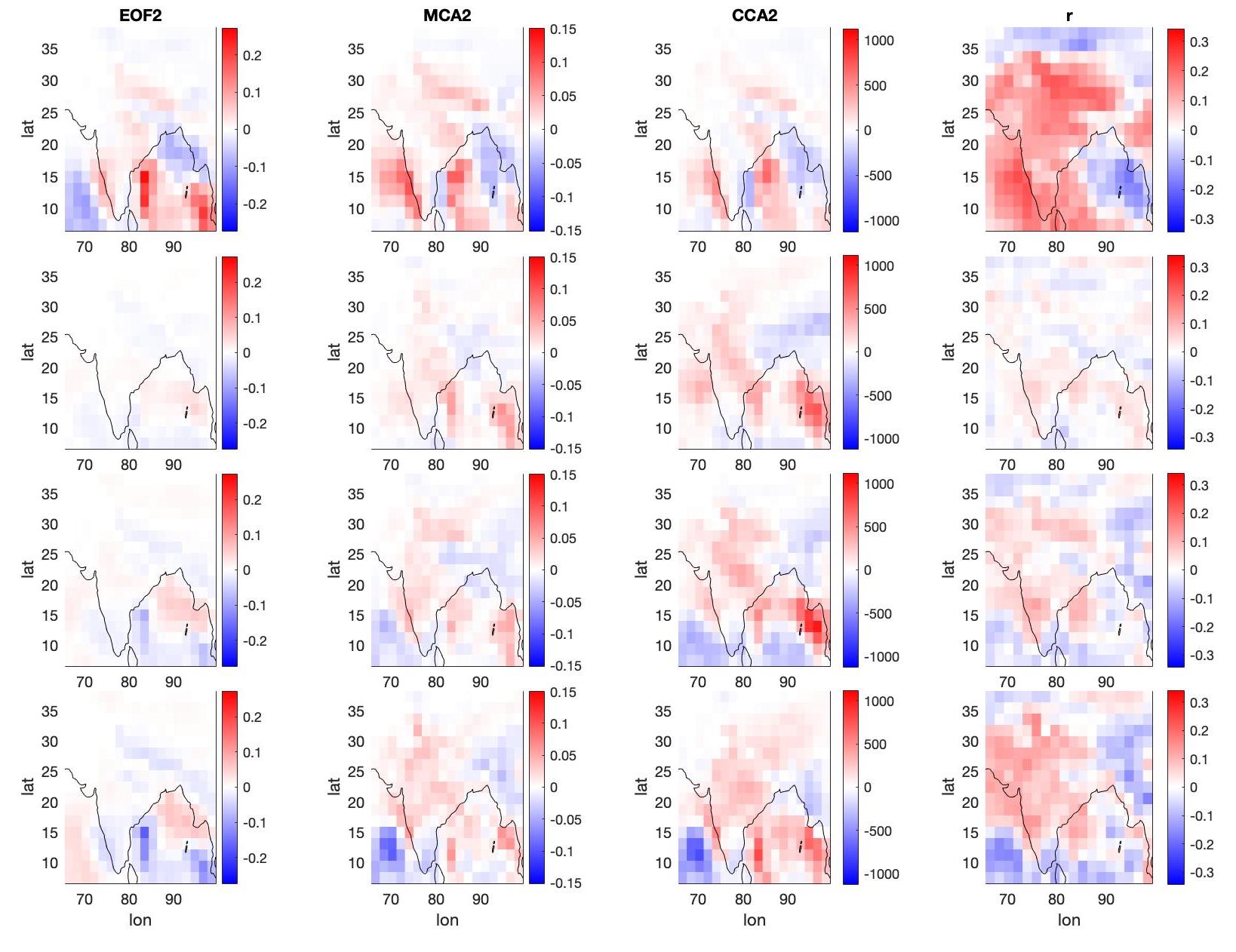} 
      \caption{\label{fig:S06} Same as Fig.~\ref{fig:S02} but the ENSO is characterised by the SLP.}
  \end{center}
\end{figure*}

\begin{figure*}
  \begin{center}
            \includegraphics[width=\linewidth]{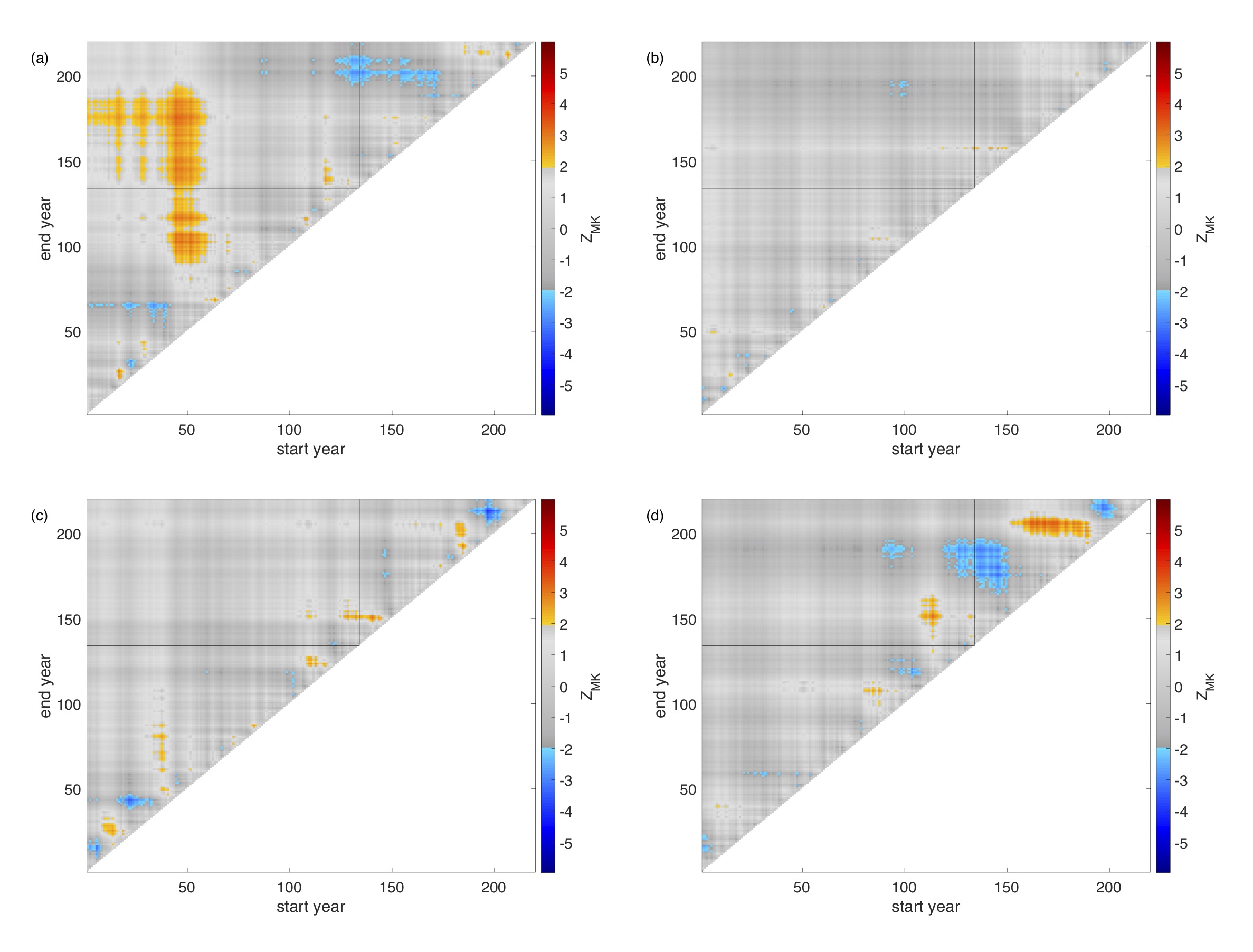}
        \caption{\label{fig:S07} Same as Fig. 4 %~\ref{fig:05} 
        (SST) but for the next-to-dominant modes.}
    \end{center}
\end{figure*}

% \begin{figure*}
%   \begin{center}
%         \begin{tabular}{cc}
%             %\hspace{0.5\linewidth} 
%             \includegraphics[width=0.5\linewidth]{ENSO_IM_teleco_14_contrib_ZMK_r_eof2_e.jpg}
%             \includegraphics[width=0.5\linewidth]{ENSO_IM_teleco_14_contrib_ZMK_r_eof22_e.jpg} \\
%             \includegraphics[width=0.5\linewidth]{ENSO_IM_teleco_14_contrib_ZMK_r_mca_2.jpg} 
%             \includegraphics[width=0.5\linewidth]{ENSO_IM_teleco_14_contrib_ZMK_r_cca_2.jpg}
%         \end{tabular}
%         \caption{\label{fig:S08} Same as Fig. 7 %~\ref{fig:06} 
%         (SLP) but for the next-to-dominant modes.}
%     \end{center}
% \end{figure*}

\begin{figure*}
  \begin{center}
            \includegraphics[width=\linewidth]{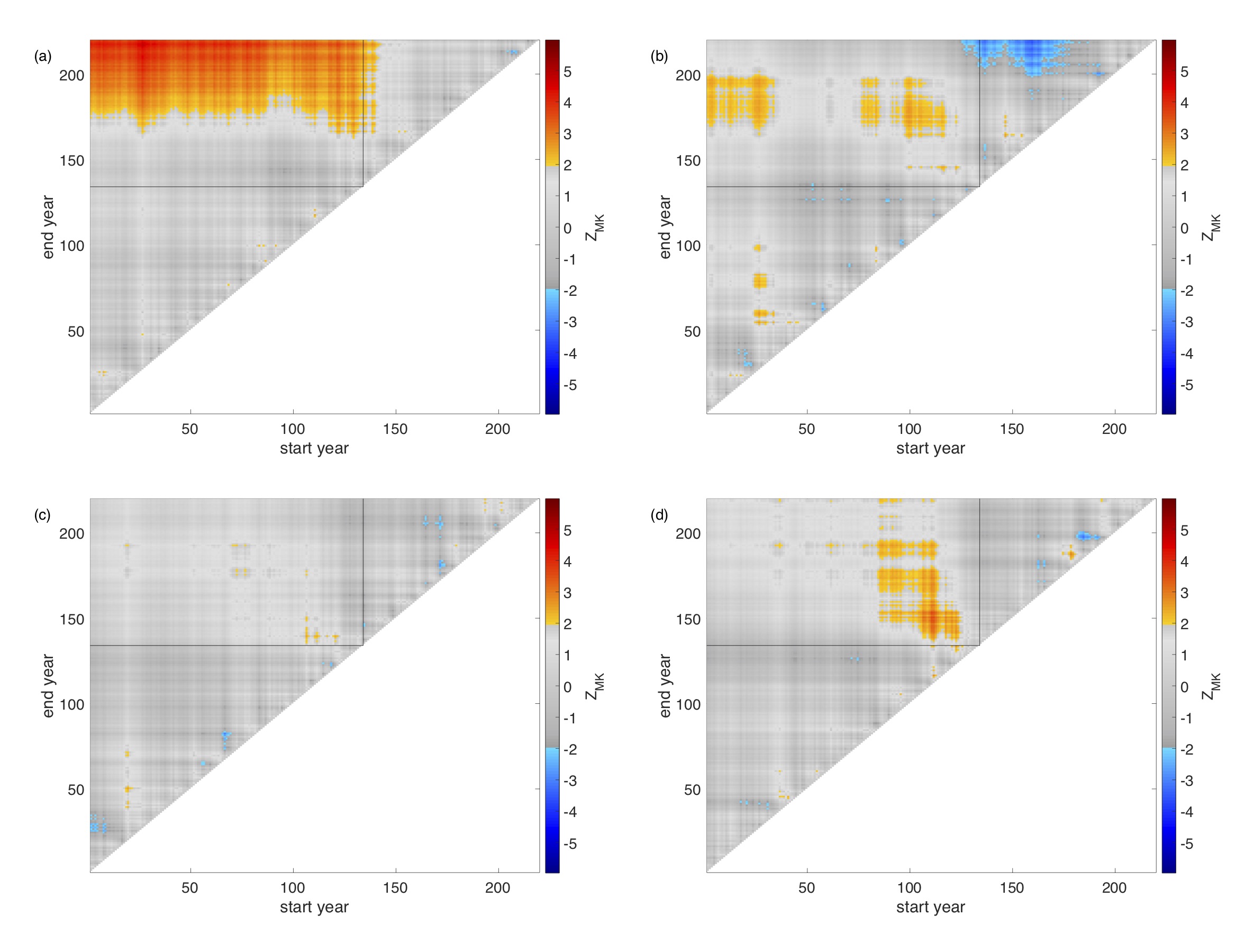}
        \caption{\label{fig:S08} Same as Fig. 7 %~\ref{fig:06} 
        (SLP) but for the next-to-dominant modes.}
    \end{center}
\end{figure*}

%\section{Nonergodicity}

\begin{figure*}
  \begin{center}
%         \begin{tabular}{cc}
%             \includegraphics[width=0.5\linewidth]{EVS_T_grid_vs_covar_01_01.jpg} 
%             \includegraphics[width=0.5\linewidth]{EVS_T_grid_vs_covar_01_04.jpg} 
%         \end{tabular}
        %\includegraphics[width=\linewidth]{ENSO_IM_teleco_19_sqmap}
	\includegraphics[width=\linewidth]{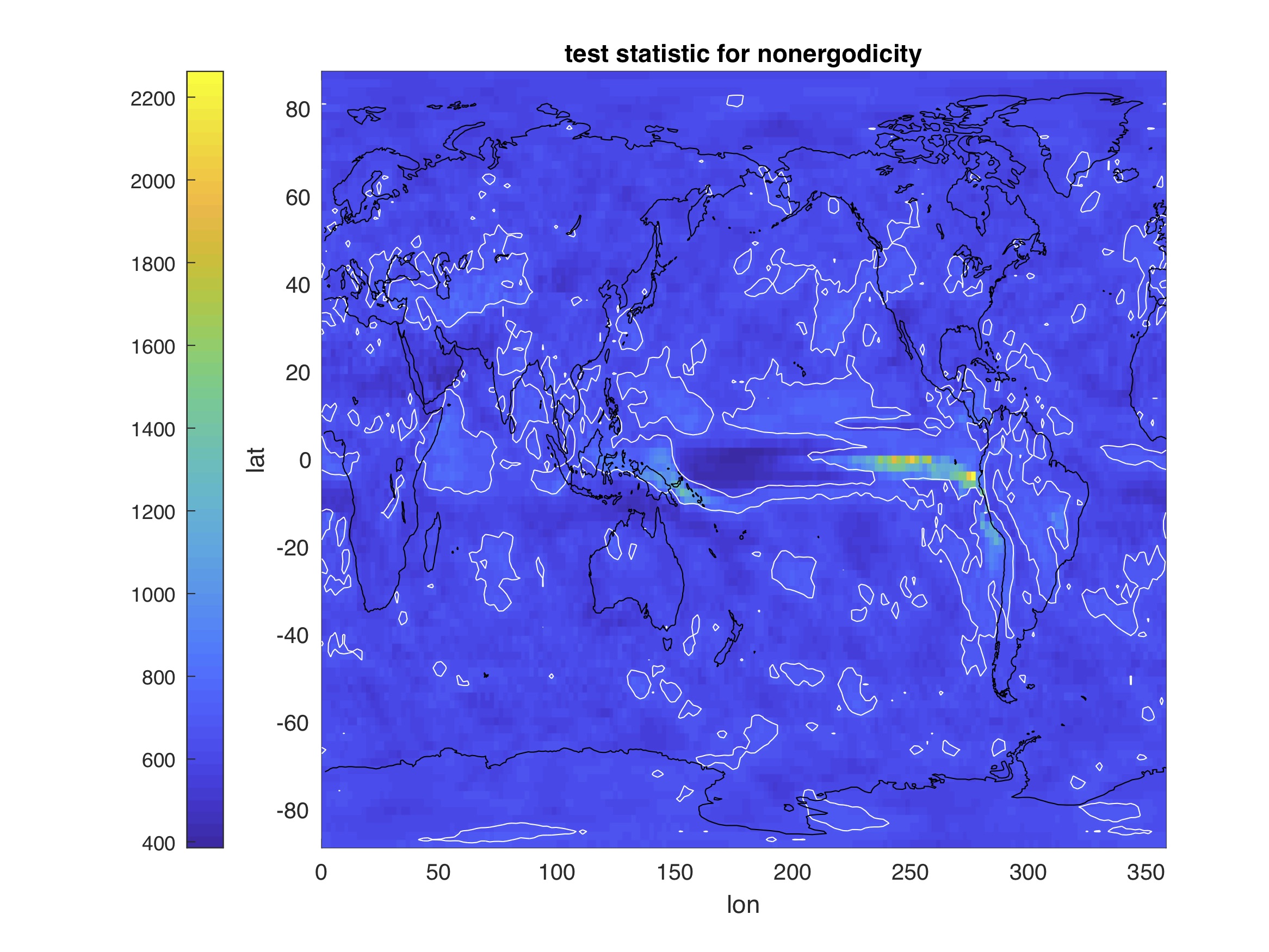}
        \caption{\label{fig:S10} Same as Fig. 8 %~\ref{fig:07} 
        but eliminating the contribution from the ensemble-mean changes.}
    \end{center}
\end{figure*}

\end{document}